\documentclass[times,usenatbib]{mn2e}
\RequirePackage{graphicx}
\RequirePackage{subfigure}
\RequirePackage{eqnarray}

\title[A suggested evolutionary diagram for prestellar cores]{The initial conditions of isolated star formation -- X. A suggested evolutionary diagram for prestellar cores}

\author[Simpson et al.]{R. J. Simpson$^{1,2}$, D. Johnstone$^{3,4}$, D. Nutter$^{1}$, D. Ward-Thompson$^{1}$, A. P. Whitworth$^{1}$ \\
$^{1}$Department of Physics and Astronomy, Cardiff University, Queen's Buildings, The Parade, Cardiff, CF24 3AA, U.K. \\
$^{2}$Oxford Astrophysics, Denys Wilkinson Building, Keble Road, Oxford, OX1 3RH, U.K. \\
$^{3}$National Research Council of Canada, Herzberg Institute of Astrophysics, 5071 West Saanich Road, Victoria, BC V9E 2E7, Canada \\
$^{4}$Department of Physics and Astronomy, University of Victoria, Elliot Building, 3800 Finnerty Road, Victoria, BC V8P 5C2, Canada }

\date{Accepted 20101 May; received 2010 November; in original form 2010 November.}

\pagerange{000--000}

\begin{document}

\label{firstpage}

\maketitle

\begin{abstract}
We propose an evolutionary path for prestellar cores on the radius-mass diagram, which is analogous to stellar evolutionary paths on the Hertzsprung-Russell Diagram. Using James Clerk Maxwell Telescope (JCMT) observations of L1688 in the Ophiuchus star-forming complex, we analyse the HCO$^{+}$ (J=4$\rightarrow$3) spectral line profiles of prestellar cores. We find that of the 58 cores observed, 14 show signs of infall in the form of a blue-asymmetric double-peaked line profile. These 14 cores all lie beyond the Jeans mass line for the region on a radius-mass plot. Furthermore another 10 cores showing tentative signs of infall, in their spectral line profile shapes, appear on or just over the Jeans mass line. We therefore propose the manner in which a prestellar core evolves across this diagram. We hypothesise that a core is formed in the low-mass, low-radius region of the plot. It then accretes quasi-statically, increasing in both mass and radius. When it crosses the limit of gravitational instability it begins to collapse, decreasing in radius, towards the region of the diagram where protostellar cores are seen.
\end{abstract}

\section{Introduction}

Star formation in molecular clouds occurs within pre-protostellar (or prestellar for short) cores, which are gravitationally bound starless cores within the clouds \citep{1994MNRAS.268..276W, 1996A&A...314..625A, 1999MNRAS.305..143W, 2002Sci...295...76W, 2007prpl.conf...33W}.

Prestellar cores represent the initial conditions for protostellar collapse. Their large-scale physical and chemical properties have been extensively studied. See, for example, \citet{1990ApJ...365..269L, 1998A&A...336..150M, 2007A&A...472..519A} and also \citet[][and references therein]{2007prpl.conf...17D}. However, the manner in which they evolve is still a matter of some debate \citep[][and references therein]{2007prpl.conf...33W}. In this paper we attempt to address the subject of prestellar core evolution and quantify this evolution with an empirical diagram analogous to the Hertzsprung-Russell Diagram.

In this series of papers we have been investigating the properties of isolated prestellar cores. The term 'isolated' refers to the idea that these cores are not interacting with each other \citep{2007A&A...472..519A}. Paper I \citep{1994MNRAS.268..276W} marked the discovery of prestellar cores. Paper II \citep{1996A&A...314..625A} and Paper III \citep{1999MNRAS.305..143W}, probed 1.3~mm continuum emission from nine prestellar cores with the IRAM telescope and studied their density profiles. Papers IV -- VII \citep{2001MNRAS.323.1025J, 2002MNRAS.329..257W, 2005MNRAS.360.1506K, 2007MNRAS.375..843K} used C$^{18}$O spectral line, ISO, SCUBA and Spitzer observations respectively to determine the density and temperature profiles of prestellar cores as well as estimating other properties such as their masses and lifetimes. Paper VIII \citep{2008MNRAS.391..205S} used SCUBA data to measure the masses and radii of prestellar cores in L1688 to demonstrate that the stellar Initial Mass Function (IMF) has its origins in the Core Mass Function (CMF). Paper IX \citep{2009MNRAS.396.1851N} studied prestellar cores in the far-infrared, using Akari observations, and demonstrated that such observations could be dominated by temperature gradients. In this paper we look at infall in prestellar cores.

\begin{table*}
\caption{Table of regions showing date of observation, position (in Right Ascension, Declination), size (in arcseconds), position angle of observed HARP maps in this investigation. Right Ascension and Declination of map off positions also shown. All dates of observations are from 2009.}
\begin{center}
\begin{tabular}{|ccccccccc}
\hline
Map & Dates of & R.A. & Declination & Width & Height & PA & Off Pos. R.A. & Off Pos. Declination \\
& Observations & (2000) & (2000) & ($^{\prime\prime}$) & ($^{\prime\prime}$) & ($^{\circ}$) & (2000) & (2000) \\
\hline
Oph-A & 25-26 Apr, 01 May & 16$^{\rm h}$26$^{\rm m}$24.03$^{\rm s}$ & $-$24$^{\circ}$23$^{\prime}$43.0$^{\prime\prime}$ & 690 & 349 & 45 & 16$^{\rm h}$25$^{\rm m}$52.99$^{\rm s}$ & $-$24$^{\circ}$25$^{\prime}$23.0$^{\prime\prime}$ \\
Oph-AN & 12,20 Jun & 16$^{\rm h}$26$^{\rm m}$44.07$^{\rm s}$ & $-$24$^{\circ}$18$^{\prime}$53.7$^{\prime\prime}$ & 420 & 348 & 50 & 16$^{\rm h}$27$^{\rm m}$05.03$^{\rm s}$ & $-$24$^{\circ}$24$^{\prime}$58.3$^{\prime\prime}$ \\
Oph-B & 15,17,22 May, 02 Jun & 16$^{\rm h}$27$^{\rm m}$22.93$^{\rm s}$ & $-$24$^{\circ}$28$^{\prime}$17.9$^{\prime\prime}$ & 550 & 348 & 150 & 16$^{\rm h}$27$^{\rm m}$05.03$^{\rm s}$ & $-$24$^{\circ}$24$^{\prime}$58.3$^{\prime\prime}$ \\
Oph-C & 08,10-11 Jun & 16$^{\rm h}$26$^{\rm m}$55.78$^{\rm s}$ & $-$24$^{\circ}$34$^{\prime}$16.5$^{\prime\prime}$ & 420 & 348 & 50 & 16$^{\rm h}$27$^{\rm m}$05.03$^{\rm s}$ & $-$24$^{\circ}$24$^{\prime}$58.3$^{\prime\prime}$ \\
Oph-E,F & 03,04,06-08 Jun & 16$^{\rm h}$27$^{\rm m}$15.76$^{\rm s}$ & $-$24$^{\circ}$39$^{\prime}$52.4$^{\prime\prime}$ & 600 & 348 & 40 & 16$^{\rm h}$27$^{\rm m}$07.82$^{\rm s}$ & $-$24$^{\circ}$44$^{\prime}$14.3$^{\prime\prime}$ \\
\hline
\end{tabular}
\end{center}
\label{maps}
\end{table*}

The Ophiuchus star-forming region is located at a distance of 139$\pm$6~pc \citep{2008AN....329...10M} and is a site of low-mass star formation \citep{1983ApJ...274..698W}.The region consists of two main clouds, L1688 and L1689, which have extended streamers leading out to distances of around 10 pc \citep{1989ApJ...338..902L}. Specifically, it is the more massive of the two clouds, L1688, that is studied in this paper, and that is generally known as the Oph main cloud. Very high star formation rates have been measured here with 14-40\% of the molecular gas being converted into stars \citep{1977AJ.....82..198V}.

The distance to Ophiuchus is a matter of recent debate. \citet{2008ApJ...675L..29L} use radio-emission  parallax observations from two young stars in Ophiuchus to calculate a preliminary distance measurement of 120$^{+ 4.5}_{- 4.2}$~pc. \citet{2008AN....329...10M} and \citet{2008A&A...480..785L} combine extinction maps with parallaxes from Hipparcos and Tycho. These studies find distances of 139$\pm$6~pc and 119$\pm$6~pc respectively. For this analysis we adopt the \citet{2008AN....329...10M} value because it uses direct parallax measurements of a larger number of objects known to be associated with L1688.

The Oph main cloud has been observed in many wavelengths from the visible to the submillimetre \citep{1983ApJ...269..182M,1989ApJ...340..823W,1990ApJ...365..269L,1992ApJ...401..667A,1992ApJ...395..516G,1997ApJS..112..109B, 1998A&A...336..150M, 2000ApJ...545..327J, 2001MNRAS.323.1025J, 2004ApJ...611L..45J, 2006A&A...447..609S, 2008ApJ...684.1240E, 2008ApJ...683..822J, 2008ApJ...672.1013P, 2009ApJ...697.1457F}. Because of this, its properties (e.g. mass, density, temperature, distance) are very well known and it is therefore a good place to study prestellar cores.

\citet{2008MNRAS.391..205S} used SCUBA archive data to produce the deepest 850-$\mu$m map of L1688. This was used to determine the masses and radii of the prestellar cores in the cloud and to produce a core mass function (CMF). It was shown that the CMF can be fitted to a three-part power law consistent with the form of the stellar initial mass function (IMF). The form of the IMF was hypothesised to be a direct result of the CMF. A similar result was seen for the Orion star-forming region by \citet{2007MNRAS.374.1413N}.

In this paper we present HCO$^{+}$ (J=4$\rightarrow$3) observations of L1688 taken using the Heterodyne Array Receiver Programme (HARP) instrument on the James Clerk Maxwell Telescope (JCMT). HCO$^{+}$ (J=4$\rightarrow$3) was chosen because it is partially optically thick, which is a requirement for any line to show the characteristic profile for infall described in Section\ref{bad-peak-description}. In addition, HCO$^{+}$ (J=4$\rightarrow$3) can be observed at high spectral and spatial resolution with the JCMT.

\section{Observations}

HARP is a spectral-imaging receiver for the JCMT, operating at submillimetre wavelengths, in the frequency range 325--375~GHz \citep{2003SPIE.4855..338S,2008SPIE.7020E..24S}. The HCO$^{+}$ (J=4$\rightarrow$3) line has a rest frequency of 356.7~GHz. HARP works in conjunction with ACSIS -- Auto-Correlation Spectral Imaging System \citep{2000ASPC..217...33D} -- a back-end correlator. It provides a spectral resolution of 0.026 kms$^{-1}$, and a spatial resolution of 14 arcseconds.

The observations here consist of 40 hours of data taken on multiple nights from April to June 2009. Five fields in L1688 were mapped, and these are outlined in Table~\ref{maps}. The maps were taken in raster position-switch mode. In this mode spectra are taken on the fly, with the telescope scanning in a direction parallel to the sides of the map. At the end of each scan row, the array is displaced by one array-width perpendicular to the scan direction. Another map is created by scanning in a perpendicular direction to the first. These can be combined in a `basket-weave' pattern to create a map with minimal striping due to differences in receptor noise. Separate fixed off-positions were used for each region. These are given in Table~\ref{maps}. These positions were selected as having no 850$\mu$m emission \citep{2008MNRAS.391..205S}. They were also checked for HCO$^{+}$ (J=4$\rightarrow$3) emission using a 60-second position-switched `stare' observation.

\begin{figure*}
\includegraphics[width=0.80\textwidth]{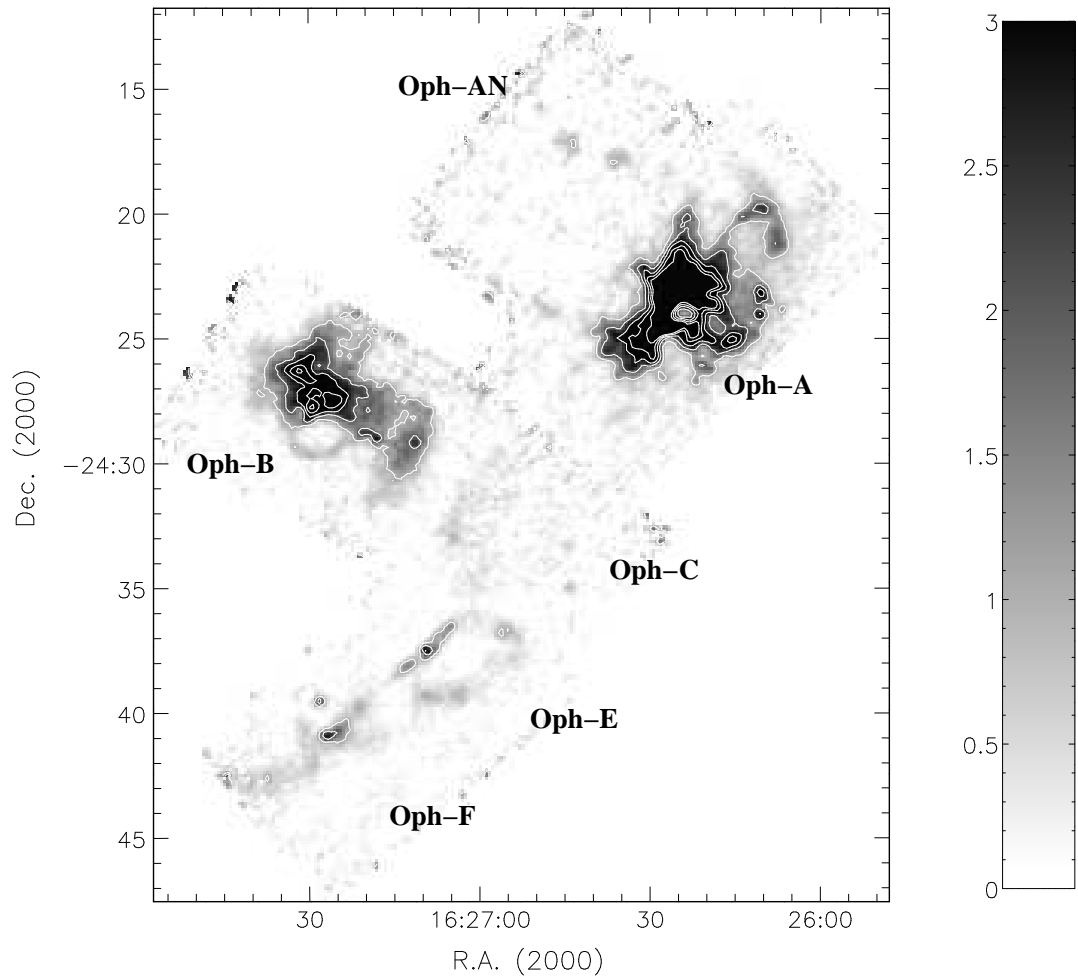}
\caption{Integrated intensity map of HCO$^{+}$ (J=4$\rightarrow$3) toward L1688. Contours at 1.0, 2.0, 3.0, 4.0 and 5.0 Kkms$^{-1}$. The scale bar shows integrated intensity in Kkms$^{-1}$. All temperatures refer to antenna temperature (T$_{A}^{*}$).}
\label{intmap}
\end{figure*}

The data were reduced using the Starlink \footnote{http://starlink.jach.hawaii.edu/starlink} project software KAPPA \citep[][and references therein]{2008ASPC..394..650C}  and SMURF \citep{2008ASPC..394..565J} routines. The data were regridded to a Nyquist-sampled map. It was found that in order to produce the optimal map from the input data, the worst quality data had to be discarded before the map was reconstructed. This was done by measuring the root-mean-square (RMS) noise value of each of the spectra in the input data and masking those with the highest RMS before generating the map. An iterative method was used to determine the optimal RMS value to apply as a mask. In each step of the iteration, the data were masked using a different limiting RMS before being reconstructed into a map using the MAKECUBE routine \citep{2008ASPC..394..565J}. The RMS in the reconstructed map was then measured. The limiting RMS was varied to produce a reconstructed map with the lowest RMS. This map was taken to be the optimal map.

\section{Results}

\subsection{Maps}

Figure~\ref{intmap} shows the integrated intensity map towards L1688 in HCO$^{+}$ (J=4$\rightarrow$3). A number of sources are seen in the map and the correlation between this integrated HCO$^{+}$ (J=4$\rightarrow$3) map and the submillimetre continuum map shown in \citet{2008MNRAS.391..205S} is consistent, showing that we are tracing the same material. The main regions of the cloud are labelled according to their usual names \citep{1998A&A...336..150M}.

\begin{figure}
\includegraphics[width=0.49\textwidth]{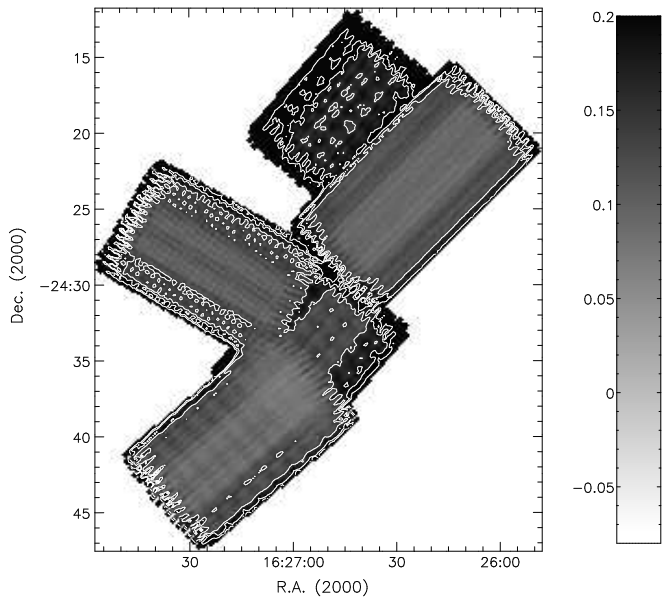}
\caption{HCO$^{+}$ (J=4$\rightarrow$3) noise map for L1688. Contours at 0.15, 0.20 and 0.25 K. The scale bar shown is also in units of K. All temperatures refer to antenna temperature (T$_{A}^{*}$).}
\label{rmsmap}
\end{figure}

The RMS noise map for the HCO$^{+}$ (J=4$\rightarrow$3) data is shown in Figure~\ref{rmsmap}. Excellent weather conditions gave a very low RMS noise of less than 0.15~K in antenna temperature, T$_{A}^{*}$, for maps of Oph-A, E and F. The noise in Oph-B increases to around 0.2~K toward the edges but most cores are still within the 0.15~K contours. Poorer weather during mapping of Oph-C resulted in a more mixed final map, with some portions mapped to 0.15~K and others to 0.2~K. The lower integration times on Oph-AN mean that the RMS noise level is higher in this region.

\begin{figure*}
\includegraphics[width=0.99\textwidth]{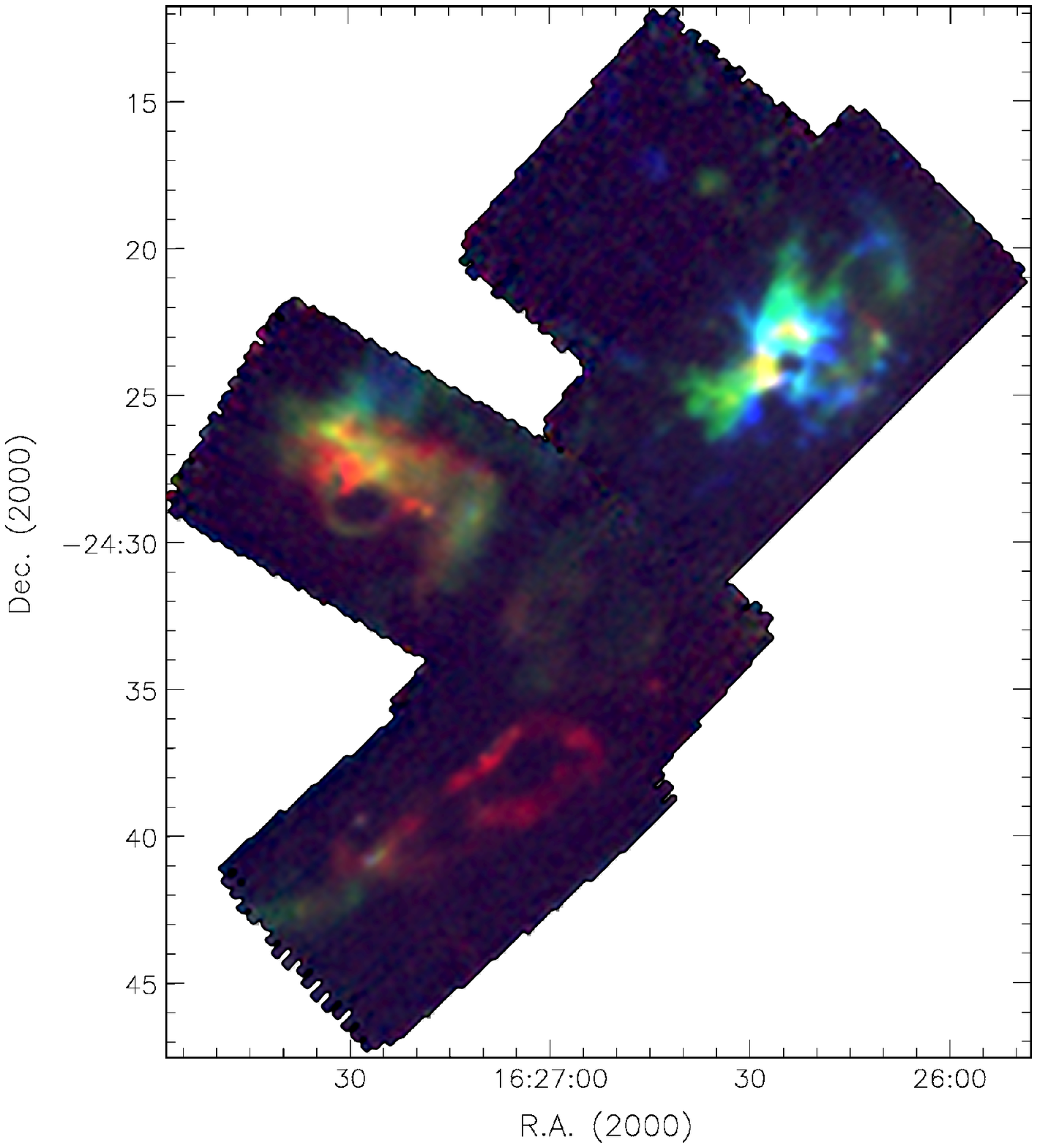}
\caption{False colour image of HCO$^{+}$ (J=4$\rightarrow$3) toward L1688 in red (4.0--5.0 kms$^{-1}$), green (3.0--4.0 kms$^{-1}$) and blue (2.0--3.0 kms$^{-1}$) velocity channels.}
\label{rgbmap}
\end{figure*}

\begin{figure*}
\begin{center}
\subfigure[4.0--5.0 kms$^{-1}$]{
\includegraphics[width=0.49\textwidth]{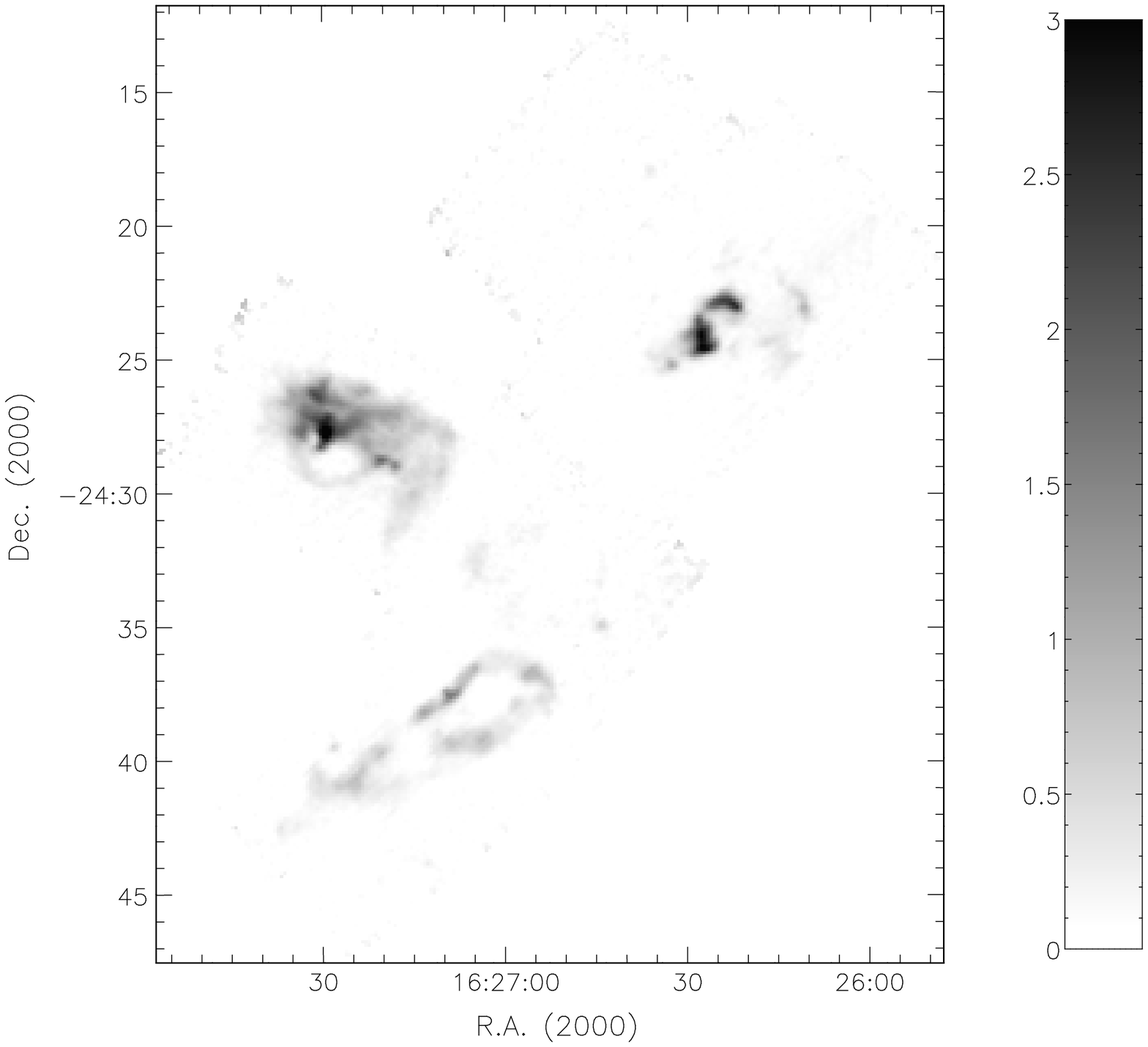}}
\subfigure[3.0--4.0 kms$^{-1}$]{
\includegraphics[width=0.49\textwidth]{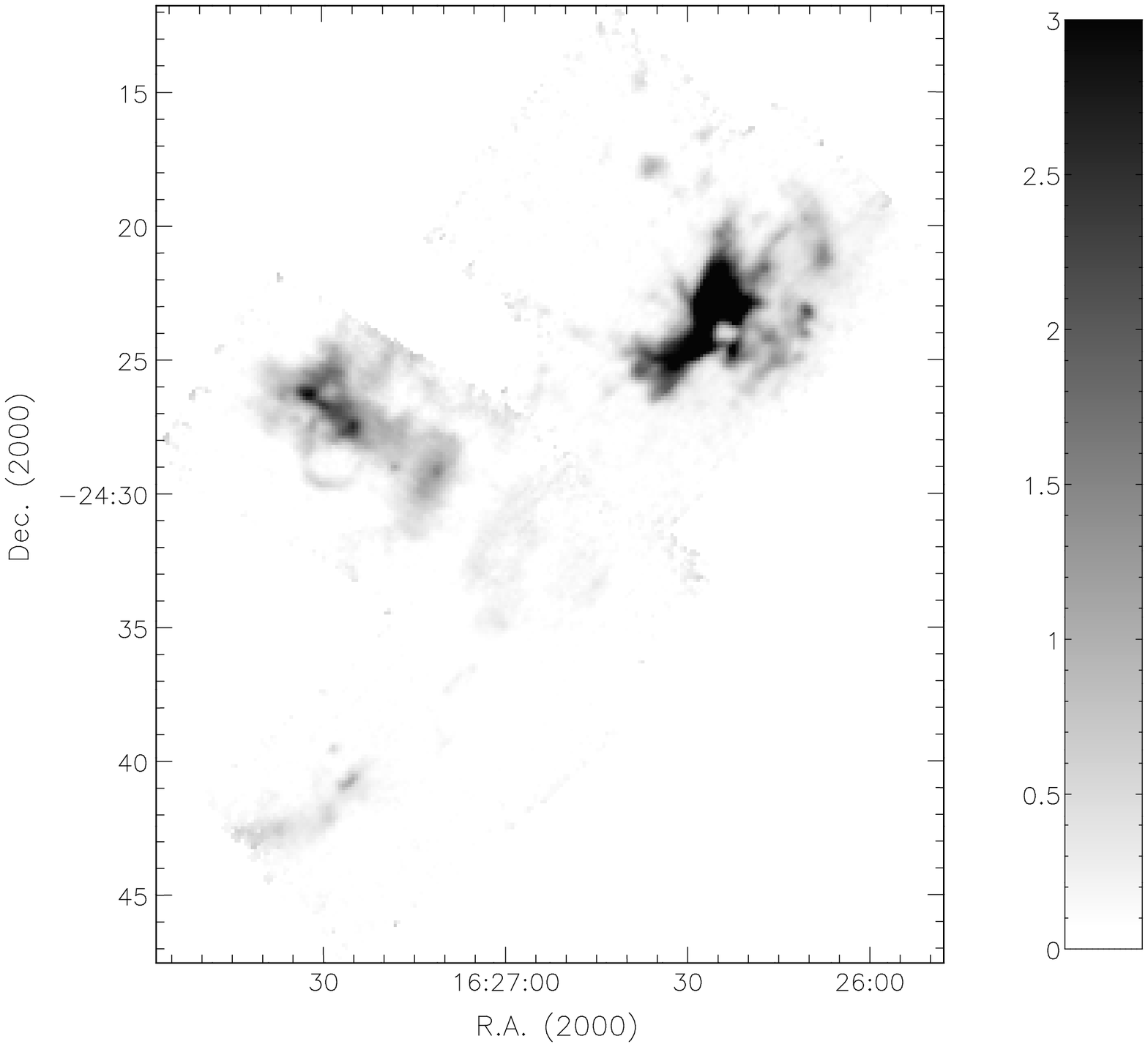}}
\subfigure[2.0--3.0 kms$^{-1}$]{
\includegraphics[width=0.49\textwidth]{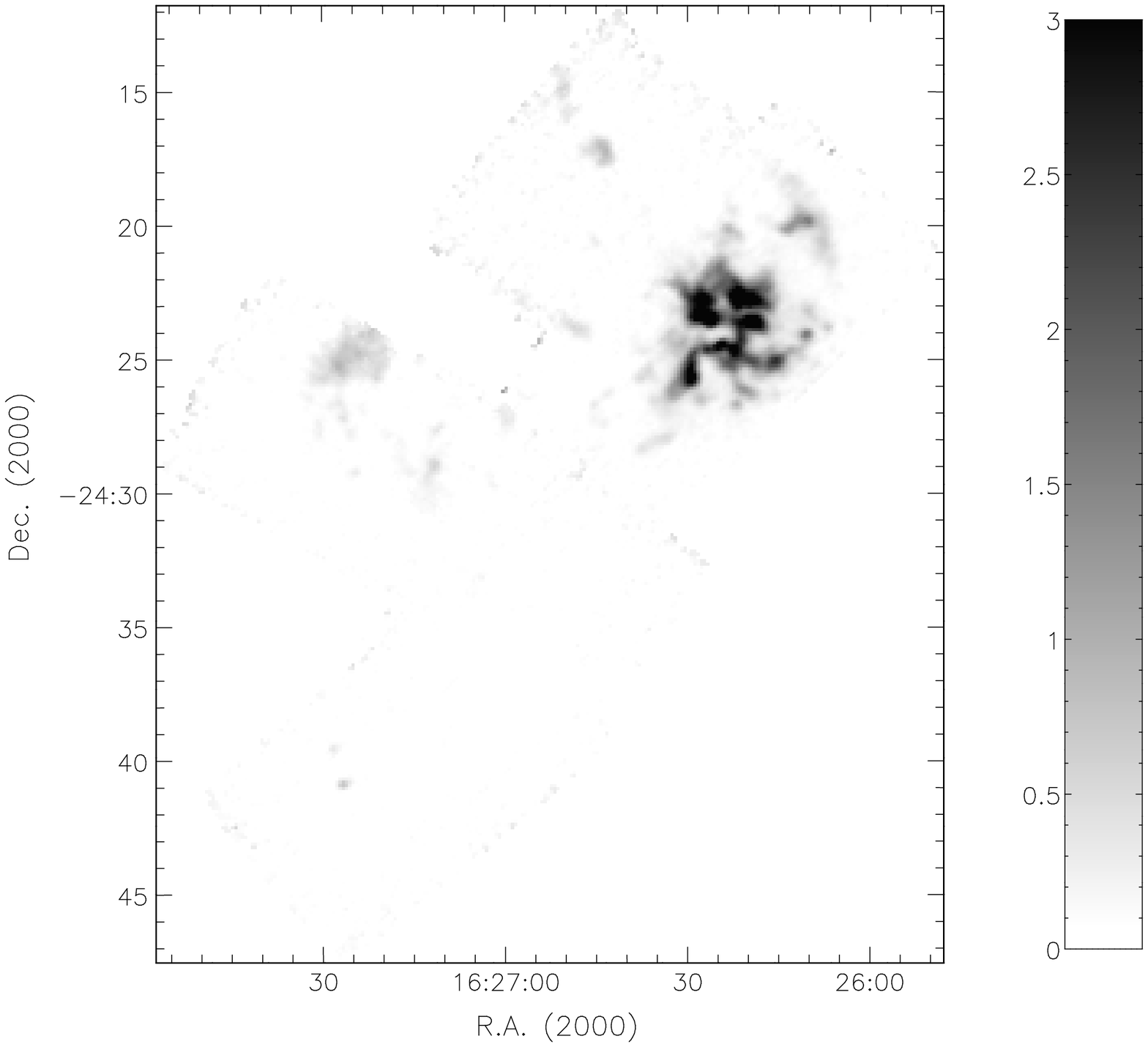}}
\caption{(a) Integrated intensity maps of HCO$^{+}$(4--3) toward L1688 in red (4.0-5.0 kms$^{-1}$), green (3.0-4.0 kms$^{-1}$) and blue (2.0-3.0 kms$^{-1}$) velocity channels. The scale bar shows integrated intensity in Kkms$^{-1}$. All temperatures refer to antenna temperature (T$_{A}^{*}$).}
\label{chanmaps}
\end{center}
\end{figure*}

Figure~\ref{rgbmap} shows the same region as Figure~\ref{intmap} but with a narrower range of velocities. This figure has been split into red (4.0--5.0 kms$^{-1}$), green (3.0--4.0 kms$^{-1}$) and blue (2.0--3.0 kms$^{-1}$)  velocity channels, centred around the systemic velocity of L1688 \citep{1989ApJ...338..902L}. These colour channels were selected to encompass the velocity widths of the observed spectral line profiles. On this version of the map it is clearly seen that Oph-A is moving toward us relative to the rest of L1688. Both Oph-A and B show evidence for HCO$^{+}$ (J=4$\rightarrow$3) outflows and a wide range of velocities are seen in both regions. Oph-E and F appear much redder and are moving away from us relative to the other regions. Figure~\ref{chanmaps} shows the same data as Figure~\ref{rgbmap} displayed as three separate images.

Emission is dominated in the integrated intensity map by the bright Oph-A region \citep[e.g.][]{2004ApJ...617..425D}, containing SM1 \citep{1989MNRAS.241..119W} and VLA1623 \citep{1993ApJ...406..122A}, respectively the prototypical prestellar core and Class 0 object. Both of these sources are highlighted by tightly packed contours in Figure~\ref{intmap}. A great deal of structure can be made out in Oph-A. An arch of material is seen traced out just northwest of the brightest part of the region -- the northern part of which is moving toward us relative to the southern part. Just south of this arch, a line of sources make up a partial loop that curves around the western edge of the Oph-A region stretching through all the velocity channels.

The Oph-B region \citep[e.g.][]{2005IAUS..235P.325F} is also very active and a distinct semi-circular loop can be seen on its south side. Both this loop and those in Oph-A have approximate radii of curvature of 0.1pc and have kinematic velocities of the order of 1-2 kms$^{-1}$. The northern part of Oph-B is seen moving toward us relative to the rest of the region. This velocity gradient was also visible in $^{13}$CO \citep{1989ApJ...338..902L}, DCO$^{+}$ \citep{1990ApJ...365..269L}, and N$_{2}$H$^{+}$ \citep{2007A&A...472..519A}.

Oph-C appears to be faint in the maps. The weak SCUBA sources measured here by \citet{2008MNRAS.391..205S} seem to have little HCO$^{+}$ (J=4$\rightarrow$3) associated with them. A large, ring-like structure is seen, in the red channel, making up the Oph-E region and this in turn leads into a filament that moves into the green channel as it joins the Oph-F region. Only the northern-most part of the Oph-F region is seen in these maps.

\subsection{Spectral Line Profiles}\label{bad-peak-description}

Prestellar cores undergoing collapse, will have excitation temperature profiles that are peaked toward their centre. In addition, we preferentially see material that is closer to us. Material on the near-side of a collapsing core is therefore preferentially seen as less excited than material on the far side. This means that the relatively blue-shifted material is seen to the observer as having a higher excitation temperature. The result is that collapsing cores display an asymmetric spectral line profile, with the peak from material at the blue-shifted end being more prominent than the red-shifted peak. Such a profile shape is referred to as a blue-asymmetric double-peaked profile and can be used as a signpost that collapse is occurring \citep[e.g.][]{1992ApJ...394..204Z, EvansII:1999gz}.

We categorise all of the observed prestellar cores based on their HCO$^{+}$ (J=4$\rightarrow$3) spectral line profile shapes: cores with blue-asymmetric double-peaked profiles; cores with faint or only possible blue-asymmetric double-peaked profiles; cores with red-asymmetric double-peaked profiles; cores with single-peaked profiles and cores with unclear profile shapes. Observations of these cores using HCO$^{+}$ (J=4$\rightarrow$3) are not sufficient to determine whether the core is undergoing collapse. Additional measurements of the cores, using an optically thin line tracer, are also required to determine that these double peaks are a spectral, rather than physical effect. For these data, we have included optically thin N$_{2}$H$^{+}$ line centres from \citet{2007A&A...472..519A}. These data were available for the majority of the observed cores and are shown on the 3$\times$3 pixel maps in the Appendix.

\begin{figure*}
\includegraphics[angle=270,width=0.90\textwidth]{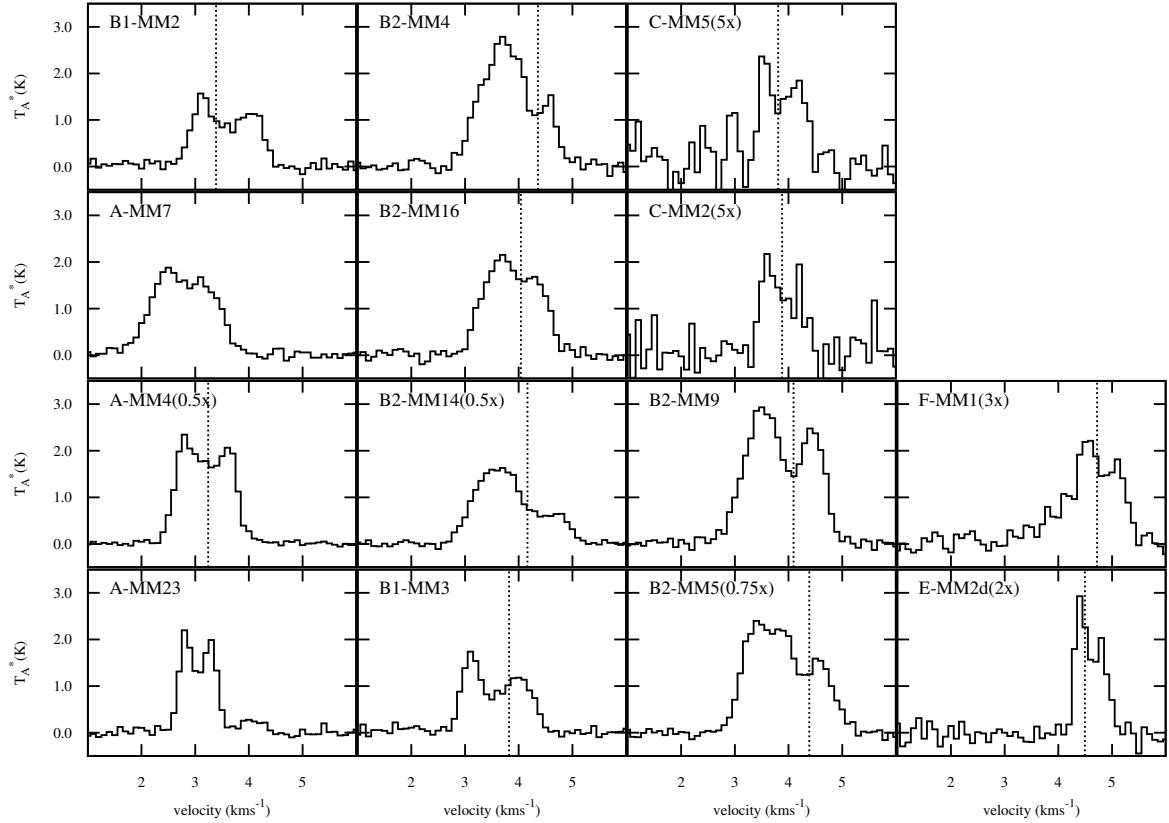}
\caption{HCO$^{+}$ (J=4$\rightarrow$3) spectral line profiles for cores mapped for this study using HARP. Cores shown here have characteristic blue-shifted, asymmetric double-peaked profiles. Systemic core velocities \citep{2007A&A...472..519A} are shown as dotted vertical lines when available.}
\label{cores_bad}
\end{figure*}

\begin{figure*}
\includegraphics[angle=270,width=0.90\textwidth]{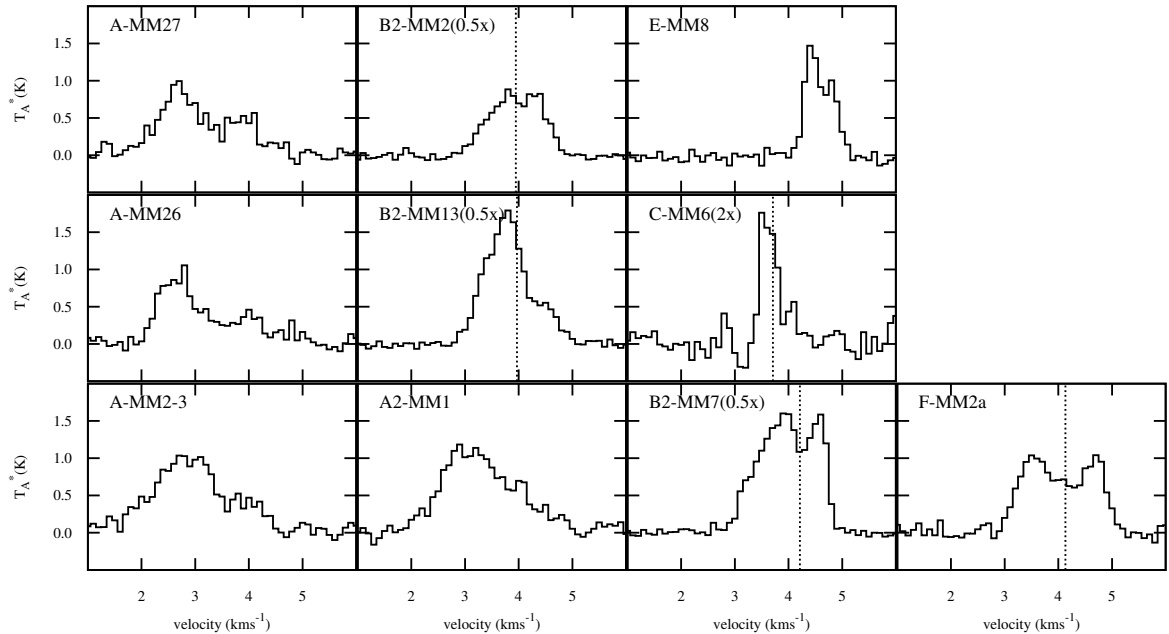}
\caption{HCO$^{+}$ (J=4$\rightarrow$3) spectral line profiles for cores mapped for this study using HARP. Cores shown here display profiles with possible blue-shifted, asymmetric double-peaks. Systemic core velocities \citep{2007A&A...472..519A} are shown as dotted vertical lines when available.}
\label{cores_badpos}
\end{figure*}

\begin{figure*}
\includegraphics[angle=270,width=0.90\textwidth]{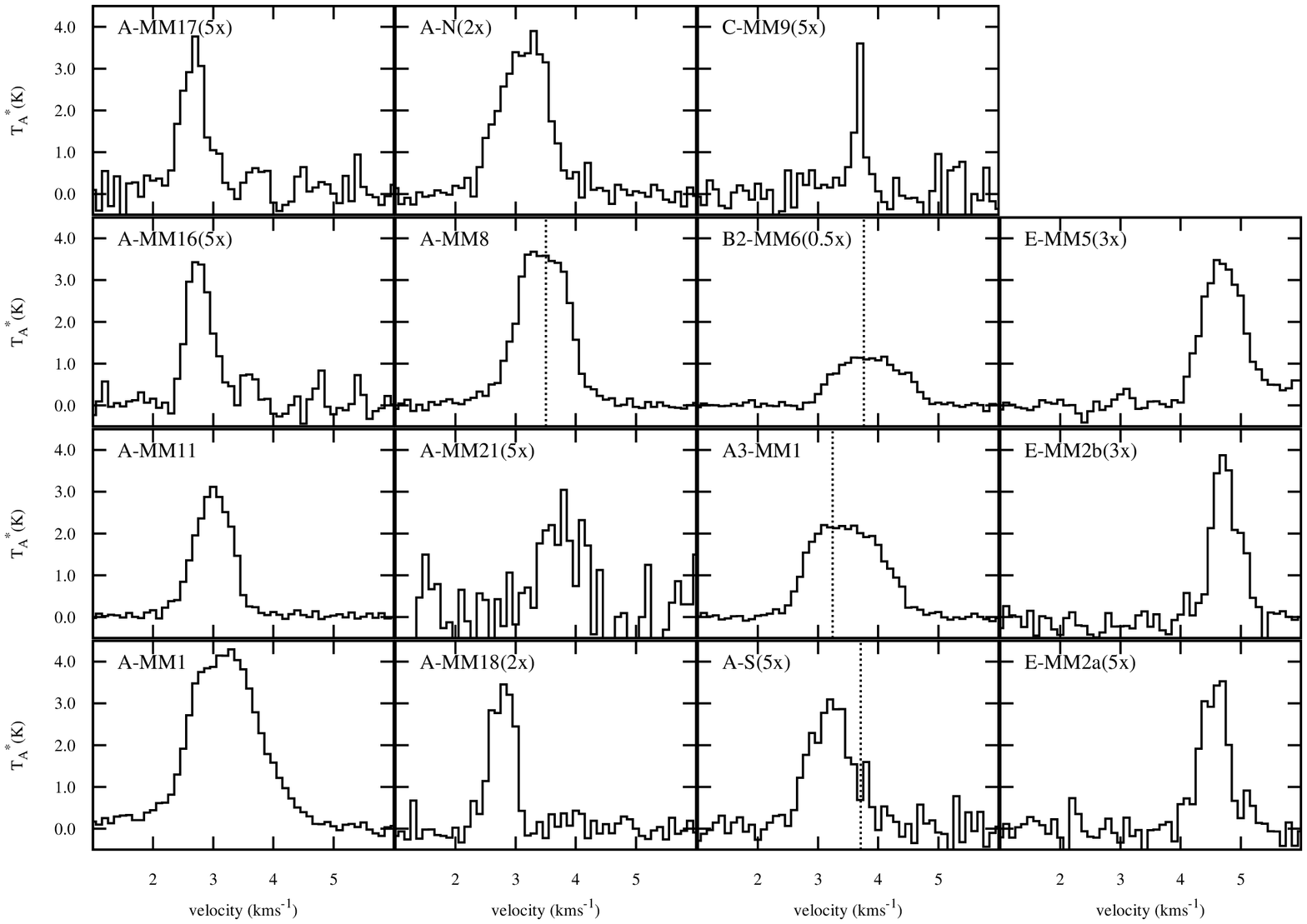}
\caption{HCO$^{+}$ (J=4$\rightarrow$3) spectral line profiles for cores mapped for this study using HARP. Cores shown here have single-peaked profiles. Systemic core velocities \citep{2007A&A...472..519A} are shown as dotted vertical lines when available.}
\label{cores_single}
\end{figure*}

\begin{figure*}
\includegraphics[angle=270,width=0.90\textwidth]{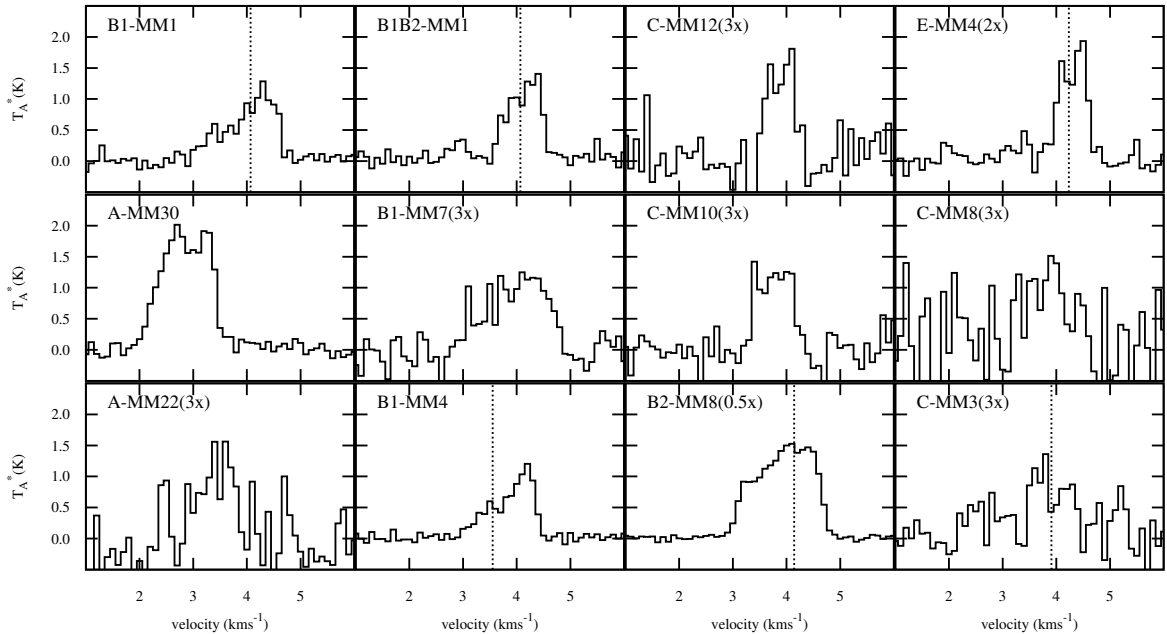}
\caption{HCO$^{+}$ (J=4$\rightarrow$3) spectral line profiles for cores mapped for this study using HARP. Cores shown here are those displaying profiles of no particular shape or low signal-to-noise. Systemic core velocities \citep{2007A&A...472..519A} are shown as dotted vertical lines when available.}
\label{cores_unclear}
\end{figure*}

Table~\ref{oph-results} shows the cores mapped in this study, grouped into regions, along with their HCO$^{+}$ (J=4$\rightarrow$3) profile shapes. The masses and radii for these cores from \citet{2008MNRAS.391..205S} are also shown. Figures~\ref{cores_bad} -- \ref{cores_rad} show the HCO$^{+}$ (J=4$\rightarrow$3) spectral line profiles for the same cores, grouped by profile shape.

\begin{table*}
\caption{Table showing the Oph cores from \citet{2008MNRAS.391..205S}, mapped in this study by HARP. For each core, the radius and mass are given along with the spectral profile shape from our HARP data. Radii are rounded to the nearest 100 AU. Blue-asymmetric and red-asymmetric double-peaked profiles are listed as BAD and RAD respectively. Whether the core was included in the \citet{1998A&A...336..150M} -- MAN98 --  is shown for reference, along with an indication of whether the core was seen as a BAD profile in \citet{2007A&A...472..519A} -- A07.}
\begin{center}
\begin{tabular}{|lccccccc}
\hline
Core & R.A. & Declination & Radius & Mass & Profile & MAN98 & A07\\
ID & (2000) & (2000) & (AU) & (M$_{\sun}$) & Shape & Core & BAD\\
\hline
A-MM1 & 16$^{\rm h}$26$^{\rm m}$23.00$^{\rm s}$ & -24$^{\circ}$23$^{\prime}$34.46$^{\prime\prime}$ & 1000 & 0.09 & Single & Y &  \\
A-MM2-3 & 16$^{\rm h}$26$^{\rm m}$23.80$^{\rm s}$ & -24$^{\circ}$24$^{\prime}$09.95$^{\prime\prime}$ & 1200 & 0.16 & BAD (Possible) & &  \\
A-MM4 & 16$^{\rm h}$26$^{\rm m}$23.03$^{\rm s}$ & -24$^{\circ}$21$^{\prime}$59.81$^{\prime\prime}$ & 1000 & 0.48 & BAD & Y &  \\
A-MM5 & 16$^{\rm h}$26$^{\rm m}$26.66$^{\rm s}$ & -24$^{\circ}$22$^{\prime}$28.60$^{\prime\prime}$ & 1300 & 0.82 & RAD & Y &  \\
A-MM6 & 16$^{\rm h}$26$^{\rm m}$28.21$^{\rm s}$ & -24$^{\circ}$23$^{\prime}$00.10$^{\prime\prime}$ & 1500 & 2.10 & RAD & Y &  \\
A-MM7 & 16$^{\rm h}$26$^{\rm m}$30.05$^{\rm s}$ & -24$^{\circ}$22$^{\prime}$17.93$^{\prime\prime}$ & 1400 & 0.92 & BAD & Y &  \\
A-MM8 & 16$^{\rm h}$26$^{\rm m}$31.80$^{\rm s}$ & -24$^{\circ}$24$^{\prime}$50.00$^{\prime\prime}$ & 1000 & 3.18 & Single & Y &  \\
A-MM11 & 16$^{\rm h}$26$^{\rm m}$32.74$^{\rm s}$ & -24$^{\circ}$26$^{\prime}$14.40$^{\prime\prime}$ & 1600 & 0.64 & Single & &  \\
A-MM16 & 16$^{\rm h}$26$^{\rm m}$36.26$^{\rm s}$ & -24$^{\circ}$28$^{\prime}$12.84$^{\prime\prime}$ & 1000 & 0.03 & Single & &  \\
A-MM17 & 16$^{\rm h}$26$^{\rm m}$34.77$^{\rm s}$ & -24$^{\circ}$28$^{\prime}$08.10$^{\prime\prime}$ & 1000 & 0.03 & Single & &  \\
A-MM18 & 16$^{\rm h}$26$^{\rm m}$43.73$^{\rm s}$ & -24$^{\circ}$17$^{\prime}$25.74$^{\prime\prime}$ & 1700 & 0.87 & Single & &  \\
A-MM21 & 16$^{\rm h}$26$^{\rm m}$31.64$^{\rm s}$ & -24$^{\circ}$18$^{\prime}$38.05$^{\prime\prime}$ & 1000 & 0.21 & Single & &  \\
A-MM22 & 16$^{\rm h}$26$^{\rm m}$31.45$^{\rm s}$ & -24$^{\circ}$18$^{\prime}$52.05$^{\prime\prime}$ & 1200 & 0.11 & Unclear & &  \\
A-MM23 & 16$^{\rm h}$26$^{\rm m}$07.89$^{\rm s}$ & -24$^{\circ}$20$^{\prime}$30.50$^{\prime\prime}$ & 1600 & 1.50 & BAD & &  \\
A-MM26 & 16$^{\rm h}$26$^{\rm m}$15.40$^{\rm s}$ & -24$^{\circ}$25$^{\prime}$32.50$^{\prime\prime}$ & 1200 & 0.43 & BAD (Possible) & &  \\
A-MM27 & 16$^{\rm h}$26$^{\rm m}$13.85$^{\rm s}$ & -24$^{\circ}$25$^{\prime}$25.16$^{\prime\prime}$ & 1400 & 0.34 & BAD (Possible) & &  \\
A-MM30 & 16$^{\rm h}$26$^{\rm m}$09.63$^{\rm s}$ & -24$^{\circ}$19$^{\prime}$43.25$^{\prime\prime}$ & 1500 & 1.28 & Unclear & &  \\
A2-MM1 & 16$^{\rm h}$26$^{\rm m}$11.73$^{\rm s}$ & -24$^{\circ}$24$^{\prime}$54.16$^{\prime\prime}$ & 1300 & 0.33 & BAD (Possible) & Y &  \\
A3-MM1 & 16$^{\rm h}$26$^{\rm m}$10.07$^{\rm s}$ & -24$^{\circ}$23$^{\prime}$11.00$^{\prime\prime}$ & 1200 & 0.33 & Single & Y &  \\
A-N & 16$^{\rm h}$26$^{\rm m}$22.74$^{\rm s}$ & -24$^{\circ}$20$^{\prime}$00.00$^{\prime\prime}$ & 1000 & 0.21 & Single & Y &  \\
A-S & 16$^{\rm h}$26$^{\rm m}$42.69$^{\rm s}$ & -24$^{\circ}$26$^{\prime}$08.05$^{\prime\prime}$ & 1000 & 0.01 & Single & &  \\
SM1 & 16$^{\rm h}$26$^{\rm m}$27.73$^{\rm s}$ & -24$^{\circ}$23$^{\prime}$58.17$^{\prime\prime}$ & 1800 & 7.35 & RAD & Y &  \\
SM1N & 16$^{\rm h}$26$^{\rm m}$27.93$^{\rm s}$ & -24$^{\circ}$23$^{\prime}$31.67$^{\prime\prime}$ & 1100 & 2.91 & Symmetric & Y &  \\
SM2 & 16$^{\rm h}$26$^{\rm m}$29.41$^{\rm s}$ & -24$^{\circ}$24$^{\prime}$26.69$^{\prime\prime}$ & 2200 & 5.97 & RAD & Y & Y \\
VLA1623 & 16$^{\rm h}$26$^{\rm m}$26.74$^{\rm s}$ & -24$^{\circ}$24$^{\prime}$30.00$^{\prime\prime}$ & 1600 & 2.93 & RAD & &  \\
\hline
B1-MM1 & 16$^{\rm h}$27$^{\rm m}$09.32$^{\rm s}$ & -24$^{\circ}$27$^{\prime}$43.73$^{\prime\prime}$ & 1100 & 0.02 & Unclear & Y &  \\
B1-MM2 & 16$^{\rm h}$27$^{\rm m}$12.14$^{\rm s}$ & -24$^{\circ}$29$^{\prime}$34.42$^{\prime\prime}$ & 1200 & 0.63 & BAD & Y &  \\
B1-MM3 & 16$^{\rm h}$27$^{\rm m}$12.68$^{\rm s}$ & -24$^{\circ}$29$^{\prime}$38.67$^{\prime\prime}$ & 1300 & 2.61 & BAD & Y &  \\
B1-MM4 & 16$^{\rm h}$27$^{\rm m}$15.32$^{\rm s}$ & -24$^{\circ}$30$^{\prime}$36.82$^{\prime\prime}$ & 2000 & 2.10 & Unclear & Y &  \\
B1-MM7 & 16$^{\rm h}$27$^{\rm m}$18.72$^{\rm s}$ & -24$^{\circ}$30$^{\prime}$24.64$^{\prime\prime}$ & 1000 & 0.11 & Unclear & &  \\
B1B2-MM1 & 16$^{\rm h}$27$^{\rm m}$12.44$^{\rm s}$ & -24$^{\circ}$27$^{\prime}$30.31$^{\prime\prime}$ & 1800 & 0.22 & Unclear & Y &  \\
B2-MM2 & 16$^{\rm h}$27$^{\rm m}$19.97$^{\rm s}$ & -24$^{\circ}$27$^{\prime}$23.45$^{\prime\prime}$ & 1400 & 0.88 & BAD (Possible) & Y &  \\
B2-MM4 & 16$^{\rm h}$27$^{\rm m}$24.50$^{\rm s}$ & -24$^{\circ}$27$^{\prime}$46.30$^{\prime\prime}$ & 1000 & 0.96 & BAD & Y &  \\
B2-MM5 & 16$^{\rm h}$27$^{\rm m}$24.74$^{\rm s}$ & -24$^{\circ}$27$^{\prime}$29.29$^{\prime\prime}$ & 1500 & 1.14 & BAD & Y &  \\
B2-MM6 & 16$^{\rm h}$27$^{\rm m}$25.27$^{\rm s}$ & -24$^{\circ}$26$^{\prime}$50.48$^{\prime\prime}$ & 2300 & 1.62 & Single & Y &  \\
B2-MM7 & 16$^{\rm h}$27$^{\rm m}$26.66$^{\rm s}$ & -24$^{\circ}$27$^{\prime}$39.72$^{\prime\prime}$ & 1700 & 0.62 & BAD (Possible) & Y &  \\
B2-MM8 & 16$^{\rm h}$27$^{\rm m}$27.96$^{\rm s}$ & -24$^{\circ}$27$^{\prime}$06.85$^{\prime\prime}$ & 1300 & 1.63 & Unclear & Y &  \\
B2-MM9 & 16$^{\rm h}$27$^{\rm m}$28.74$^{\rm s}$ & -24$^{\circ}$26$^{\prime}$40.65$^{\prime\prime}$ & 1400 & 1.37 & BAD & Y &  \\
B2-MM13 & 16$^{\rm h}$27$^{\rm m}$32.95$^{\rm s}$ & -24$^{\circ}$26$^{\prime}$03.16$^{\prime\prime}$ & 1500 & 0.49 & BAD (Possible) & Y &  \\
B2-MM14 & 16$^{\rm h}$27$^{\rm m}$32.53$^{\rm s}$ & -24$^{\circ}$26$^{\prime}$27.78$^{\prime\prime}$ & 1500 & 1.26 & BAD & Y &  \\
B2-MM16 & 16$^{\rm h}$27$^{\rm m}$34.49$^{\rm s}$ & -24$^{\circ}$26$^{\prime}$11.90$^{\prime\prime}$ & 1200 & 1.31 & BAD & Y & Y \\
\hline
C-MM2 & 16$^{\rm h}$26$^{\rm m}$59.10$^{\rm s}$ & -24$^{\circ}$33$^{\prime}$50.15$^{\prime\prime}$ & 1000 & 1.11 & BAD & Y &  \\
C-MM3 & 16$^{\rm h}$26$^{\rm m}$58.80$^{\rm s}$ & -24$^{\circ}$34$^{\prime}$23.40$^{\prime\prime}$ & 1900 & 1.42 & Unclear & Y &  \\
C-MM5 & 16$^{\rm h}$26$^{\rm m}$59.77$^{\rm s}$ & -24$^{\circ}$34$^{\prime}$13.87$^{\prime\prime}$ & 1100 & 1.24 & BAD & Y & Y \\
C-MM6 & 16$^{\rm h}$27$^{\rm m}$01.58$^{\rm s}$ & -24$^{\circ}$34$^{\prime}$44.62$^{\prime\prime}$ & 2400 & 1.05 & BAD (Possible) & Y & Y \\
C-MM8 & 16$^{\rm h}$26$^{\rm m}$49.09$^{\rm s}$ & -24$^{\circ}$29$^{\prime}$45.15$^{\prime\prime}$ & 1000 & 0.15 & Unclear & &  \\
C-MM9 & 16$^{\rm h}$26$^{\rm m}$48.10$^{\rm s}$ & -24$^{\circ}$32$^{\prime}$12.50$^{\prime\prime}$ & 1100 & 0.06 & Single & &  \\
C-MM10 & 16$^{\rm h}$27$^{\rm m}$02.26$^{\rm s}$ & -24$^{\circ}$31$^{\prime}$42.56$^{\prime\prime}$ & 1100 & 0.37 & Unclear & &  \\
C-MM12 & 16$^{\rm h}$26$^{\rm m}$59.80$^{\rm s}$ & -24$^{\circ}$33$^{\prime}$08.75$^{\prime\prime}$ & 1600 & 0.09 & Unclear & &  \\
C-N & 16$^{\rm h}$26$^{\rm m}$58.11$^{\rm s}$ & -24$^{\circ}$31$^{\prime}$46.32$^{\prime\prime}$ & 2100 & 1.60 & RAD & Y &  \\
\hline
E-MM2a & 16$^{\rm h}$27$^{\rm m}$02.89$^{\rm s}$ & -24$^{\circ}$38$^{\prime}$46.47$^{\prime\prime}$ & 1000 & 0.12 & Single & Y &  \\
E-MM2b & 16$^{\rm h}$27$^{\rm m}$02.19$^{\rm s}$ & -24$^{\circ}$39$^{\prime}$11.49$^{\prime\prime}$ & 1400 & 0.16 & Single & Y &  \\
E-MM2d & 16$^{\rm h}$27$^{\rm m}$04.53$^{\rm s}$ & -24$^{\circ}$39$^{\prime}$06.42$^{\prime\prime}$ & 1000 & 0.41 & BAD & Y & Y \\
E-MM4 & 16$^{\rm h}$27$^{\rm m}$10.67$^{\rm s}$ & -24$^{\circ}$39$^{\prime}$25.28$^{\prime\prime}$ & 1700 & 0.31 & Unclear & Y & Y \\
E-MM5 & 16$^{\rm h}$27$^{\rm m}$11.67$^{\rm s}$ & -24$^{\circ}$37$^{\prime}$57.46$^{\prime\prime}$ & 1200 & 0.28 & Single & Y &  \\
E-MM8 & 16$^{\rm h}$27$^{\rm m}$04.19$^{\rm s}$ & -24$^{\circ}$39$^{\prime}$02.59$^{\prime\prime}$ & 1200 & 0.19 & BAD (Possible) \\\hline
F-MM1 & 16$^{\rm h}$27$^{\rm m}$21.76$^{\rm s}$ & -24$^{\circ}$39$^{\prime}$48.05$^{\prime\prime}$ & 1600 & 1.05 & BAD & Y &  \\
F-MM2a & 16$^{\rm h}$27$^{\rm m}$24.43$^{\rm s}$ & -24$^{\circ}$40$^{\prime}$34.79$^{\prime\prime}$ & 2000 & 1.11 & BAD (Possible) & Y &  \\
\hline
\end{tabular}
\end{center}
\label{oph-results}
\end{table*}

\section{Discussion}

\begin{figure}
\includegraphics[angle=270,width=0.49\textwidth]{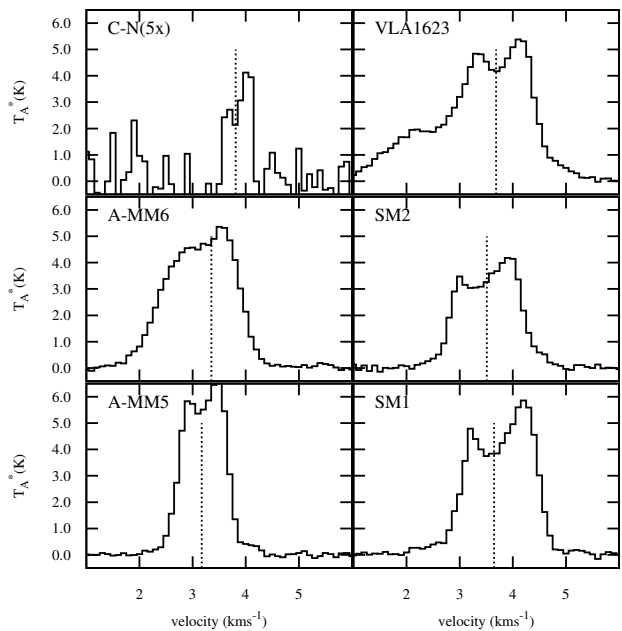}
\caption{HCO$^{+}$ (J=4$\rightarrow$3) spectral line profiles for cores mapped for this study using HARP. Cores shown here have red-asymmetric double-peaked profiles. Systemic core velocities \citep{2007A&A...472..519A} are shown as dotted vertical lines when available.}
\label{cores_rad}
\end{figure}

If we assume that the prestellar cores are at a constant gas-kinetic temperature and are thermally supported \citep{2000ApJ...545..327J, 2001ApJ...559..307J, 2007MNRAS.375..843K}, then they will have a radius-mass relationship determined by the external pressure on them. The exact solutions are given by Bonnor-Ebert spheres \citep{1955ZA.....37..217E, 1956MNRAS.116..351B}. This is in contrast to the cloud at large, where densities are lower and turbulent motions are greater \citep{2007MNRAS.375..843K}.

Assuming an approximately uniform and constant temperature for all the gas in the cloud allows us to calculate the Jeans Mass \citep{1902RSPTA.199....1J}. Below the Jeans Mass, the internal forces in the core are dominated by thermal pressure. When above the Jeans mass, the core becomes unstable and collapses under its own self-gravity.

\begin{figure*}
\includegraphics[angle=0,width=0.99\textwidth]{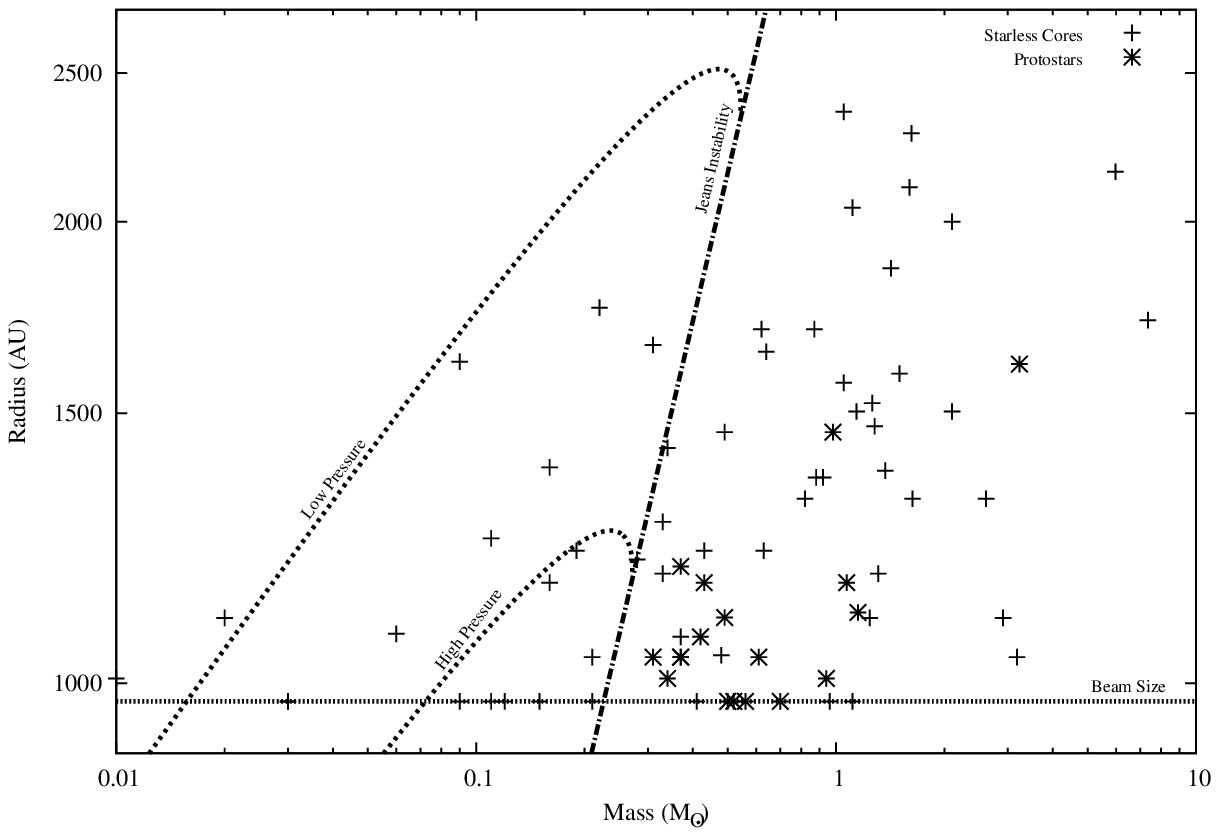}
\caption{Starless and protostellar cores from \citet{2008MNRAS.391..205S} plotted in terms of radius against mass. Starless cores are indicated with crosses. Protostellar cores (i.e. cores with infrared associations) are marked with asterisks. For error bars, see Figure~\ref{results-plot-simple}. See text for more details.}
\label{original-plot}
\end{figure*}

Figure~\ref{original-plot} shows the cores detected by \citet{2008MNRAS.391..205S} at 850$\mu$m with SCUBA, plotted in terms of radius versus mass. Here the radius is defined using the 3$\sigma$ contour of the signal-to-noise map of the region -- see \citet{2007MNRAS.374.1413N} and \cite{2007MNRAS.374.1413N}. The crosses mark the prestellar cores. The cores with associated infrared sources identified by Spitzer, as part of the c2d survey \citep{2008ApJ...684.1240E}, are deemed to be protostellar and are marked with asterisks. We assume that these objects are Class 0 or I protostars \citep{1993ApJ...406..122A, 2009ApJ...692..973E}, meaning that they have accreted only part of their final mass. In order to account for the protostellar mass not detected by SCUBA we use a mean IMF mass of 0.6$M_{\sun}$ \citep{2002Sci...295...82K} and thus increase the masses of these objects by 0.3$M_{\sun}$. This would be the correction needed if all of the objects lay exactly on the Class 0/I borderline, defined as the point at which a protostar has accreted half of its final main sequence mass \citep{1993ApJ...406..122A}, and hence is indicative of the correction needed.

\begin{figure*}
\includegraphics[angle=0,width=0.99\textwidth]{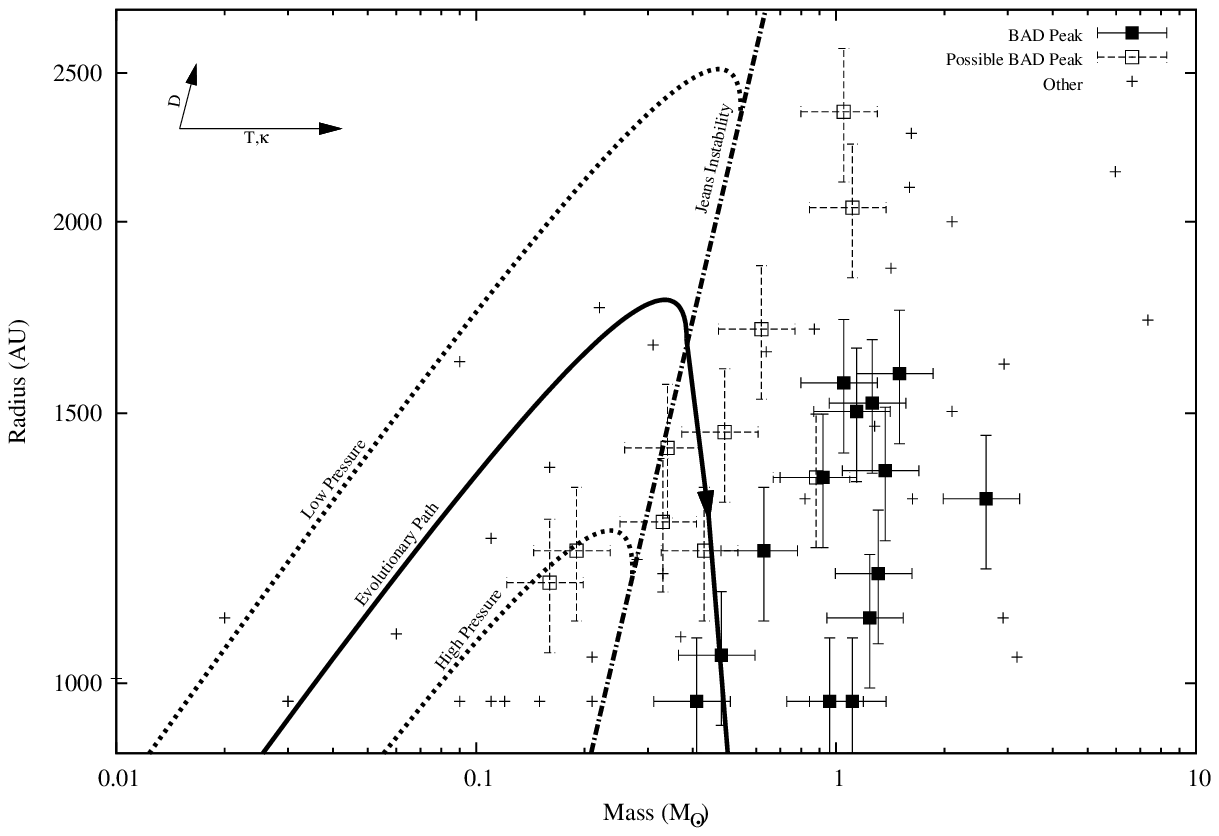}
\caption{Same as Figure~\ref{original-plot}, excluding the protostellar cores. Cores showing definite blue-asymmetric double (BAD) peaks in their spectral line profiles are shown as dark squares; cores with a possible blue-asymmetric double peak are shown as hollow squares; all other cores are shown as crosses. Systematic errors are shown in the top-left hand of the plot for distance to Ophiuchus, $D$, temperature of the cores, $T$ and dust mass opacity, $\kappa$. An error of 10\% is estimated for the measurement of the radii of cores. This produces a corresponding error in mass. These error bars are shown on the plot. We propose this as an evolutionary diagram for prestellar cores, and indicate the manner of this evolution with an arrow.}
\label{results-plot-simple}
\end{figure*}

We propose the radius versus mass diagram shown in Figure~\ref{original-plot} to be the equivalent of an HR diagram for prestellar cores. Cores evolve by accreting mass quasi-statically and maintaining Bonnor-Ebert equilibrium. Upon reaching the Jeans instability, cores collapse to form a protostar, moving downward on this diagram.

Protostellar cores lie beyond the Jeans limit, to the right on the graph, near the resolution limit of the observations. These cores have already accreted material and begun to collapse. Again these cores have been given an additional 0.3$M_{\sun}$ to account for the undetected mass.

The hypothesis that this diagram illustrates the evolution of prestellar cores can be tested by looking at the position on this diagram occupied by collapsing prestellar cores. If this evolutionary picture is correct, then the blue-asymmetric double-peaked profiles should all lie close to, or beyond the Jeans mass.

Table~\ref{oph-results} lists the masses and radii of those starless cores detected by SCUBA, that were also detected in HCO$^{+}$ (J=4$\rightarrow$3) by HARP. Their profile shapes and positions are also given. Figure~\ref{results-plot-simple} is the same as Figure~\ref{original-plot} except that it shows only the cores listed in Table~\ref{oph-results}. This time they are plotted according to their spectral line profile shapes. Those cores with blue-asymmetric profiles, indicating collapse, are marked with dark boxes. Cores showing possible infall characteristics are marked with hollow boxes. All other cores are marked as crosses. These cores may simply have the wrong physical conditions to produce a detectable blue-asymmetric profile in HCO$^{+}$ (J=4$\rightarrow$3). Figure~\ref{cores_rad} shows those cores that display red-asymmetric profiles. These are likely caused by bipolar outflows confusing the signature, especially in the case of VLA1623, which might also interfere with the neighbouring SM1 and SM2. It is noted that this explanation does not necessarily account for the profile of the core C-N, which displays a very thin RAD profile.

In Figures~\ref{original-plot} and \ref{results-plot-simple} the two dashed lines marked `Low Pressure' and `High Pressure' show the Bonnor-Ebert relations for external pressures of P/$k = 3 \times 10^{6}$ and $12 \times 10^{6}~\rm{K~cm}^{-3}$ respectively \citep{2000ApJ...545..327J, 2008ApJ...683..822J}. The solid line in Figure~\ref{results-plot-simple} shows a possible evolutionary track, indicating the general direction a core would move through the diagram. This track initially follows a Bonnor-Ebert relation for an external pressure of P/$k = 6 \times 10^{6}~\rm{K~cm}^{-3}$. We assume a temperature for the cores in Ophiuchus of 10~K \citep{2007MNRAS.379.1390S}.

The hypothesis that a radius-mass plot can serve as an evolutionary diagram for prestellar cores is supported by these results. In Figure~\ref{results-plot-simple}, cores showing only weak signs of infall (hollow boxes) appear close to, or beyond, the Jeans mass line compared with those cores exhibiting characteristic infall signatures (solid boxes) which all appear beyond it. 40\% of the cores beyond the Jeans mass line show signs of infall; 60\% if the possible BAD peaks are included. This supports the conjecture that cores follow an evolutionary track in which they first move diagonally up the diagram, as they accrete material, and then move down it as they cross the Jeans mass limit and collapse.

Using this model, for any given prestellar core on this diagram, an approximate protostellar mass could be derived by following a Bonnor-Ebert track diagonally upward to the Jeans mass line. From that point in the evolution, an efficiency for the collapse of the core needs to be assumed to reach an estimate of the final protostellar mass. If a core continues to accrete whilst collapsing, the gradient of its evolutionary track, after crossing the Jeans mass, will be shallower.

\section{Conclusions}

We propose that a radius-mass plot can serve as an evolutionary diagram for prestellar cores. Of the 58 prestellar cores observed with HARP, 14 showed signs of infall in the form of a blue-asymmetric double-peaked line profile. These 14 cores all lie beyond the Jeans mass line for the region on such a radius-mass plot. Furthermore those cores showing tentative signs of infall, in their spectral line profile shapes, appear close to, or beyond, the Jeans mass line.

Placing prestellar cores in the context of this evolutionary diagram shows how they can evolve into protostars inside molecular clouds. This analysis could be performed using data for other star-forming clouds, where both submillimetre maps and spectral profile data are available. 

Future studies using instruments such as SCUBA-2 and Herschel may be able to probe further to the left on diagram, and see how larger, lower-mass objects -- such as CO cores -- relate to the gravitationally bound prestellar cores already seen in the continuum. Such objects appear to follow different radius-mass relations to those explored in this study -- see Figure~4 of \citet{2001A&A...372L..41M} and discussions by \citet{2000ApJ...545..327J, 2001ApJ...559..307J}.

\section*{Acknowledgments}
The authors would to thank Sarah Stickler, Lucy Wilcock and the staff of the JCMT for their help whilst obtaining the data for this study. The JCMT is operated by the Joint Astronomy Centre (JAC) in Hawaii on behalf of the UK Science and Technology Facilities Council (STFC), the Netherlands Organisation for Scientific Research (NWO) and the Canadian National Research Council (NRC). RJS acknowledges STFC studentship support whilst carrying out this work. DN is supported by the Cardiff Astronomy Rolling Grant from STFC.

\bibliographystyle{mn2e}
\bibliography{library}

\clearpage

\appendix{}
\section{3x3 pixel HARP maps.}\label{app}

\begin{figure*}
\includegraphics[width=0.75\textwidth]{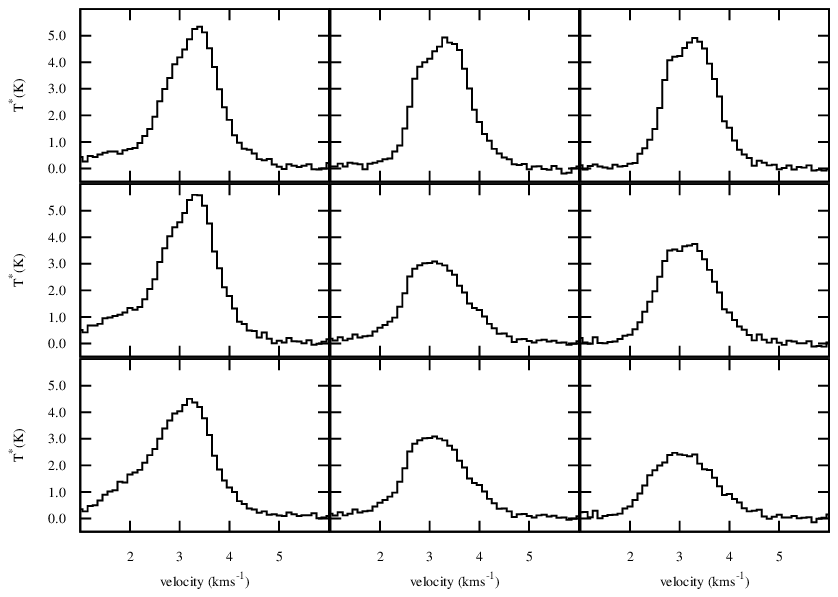}
\caption[Map of HCO$^{+}$ (J=4$\rightarrow$3) spectra in the core A-MM1.]{Map of HCO$^{+}$ (J=4$\rightarrow$3) spectra in the core A-MM1. The central grid of 3$\times$3 HARP pixels is shown here.}
\label{A-MM1}
\end{figure*}

\begin{figure*}
\includegraphics[width=0.75\textwidth]{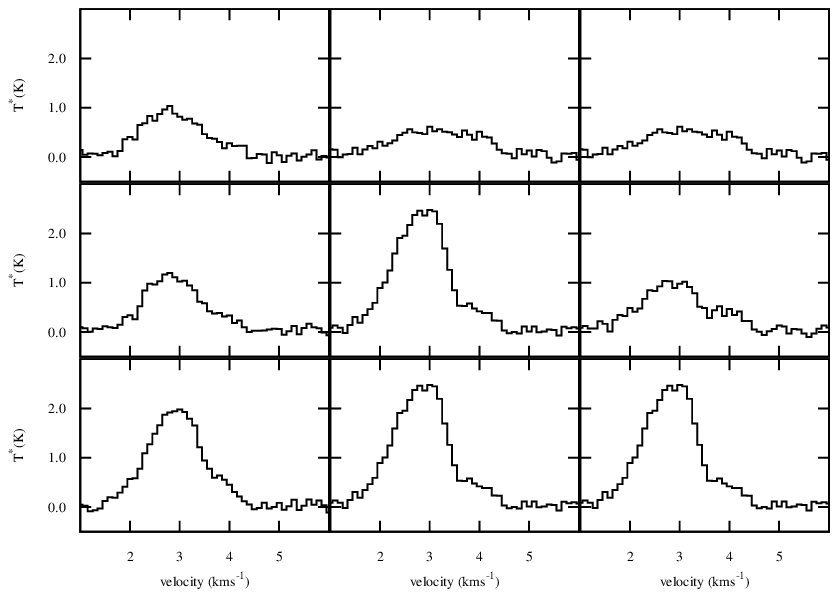}
\caption[Map of HCO$^{+}$ (J=4$\rightarrow$3) spectra in the core A-MM2-3.]{Map of HCO$^{+}$ (J=4$\rightarrow$3) spectra in the core A-MM2-3. The central grid of 3$\times$3 HARP pixels is shown here.}
\label{A-MM2-3}
\end{figure*}

\begin{figure*}
\includegraphics[width=0.75\textwidth]{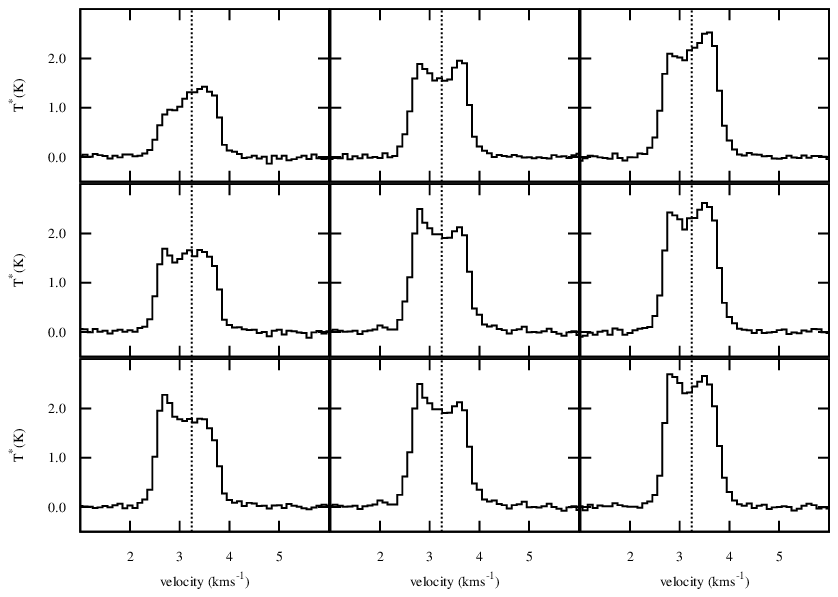}
\caption[Map of HCO$^{+}$ (J=4$\rightarrow$3) spectra in the core A-MM4.]{Map of HCO$^{+}$ (J=4$\rightarrow$3) spectra in the core A-MM4. The central grid of 3$\times$3 HARP pixels is shown here.}
\label{A-MM4}
\end{figure*}

\begin{figure*}
\includegraphics[width=0.75\textwidth]{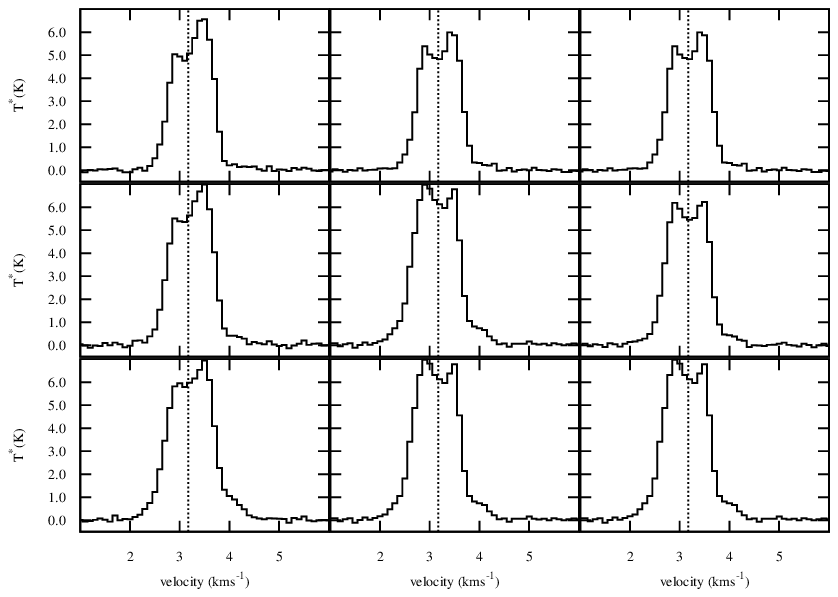}
\caption[Map of HCO$^{+}$ (J=4$\rightarrow$3) spectra in the core A-MM5.]{Map of HCO$^{+}$ (J=4$\rightarrow$3) spectra in the core A-MM5. The central grid of 3$\times$3 HARP pixels is shown here.}
\label{A-MM5}
\end{figure*}

\begin{figure*}
\includegraphics[width=0.75\textwidth]{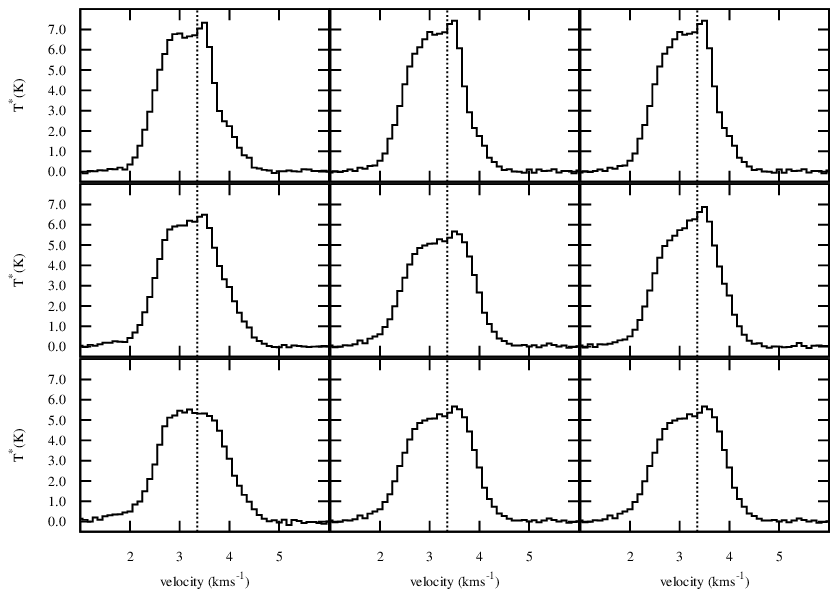}
\caption[Map of HCO$^{+}$ (J=4$\rightarrow$3) spectra in the core A-MM6.]{Map of HCO$^{+}$ (J=4$\rightarrow$3) spectra in the core A-MM6. The central grid of 3$\times$3 HARP pixels is shown here.}
\label{A-MM6}
\end{figure*}

\begin{figure*}
\includegraphics[width=0.75\textwidth]{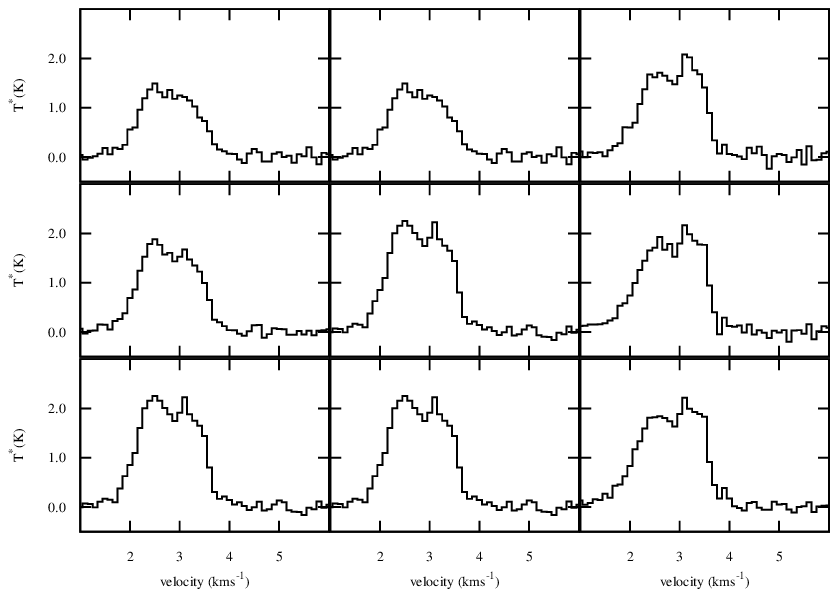}
\caption[Map of HCO$^{+}$ (J=4$\rightarrow$3) spectra in the core A-MM7.]{Map of HCO$^{+}$ (J=4$\rightarrow$3) spectra in the core A-MM7. The central grid of 3$\times$3 HARP pixels is shown here.}
\label{A-MM7}
\end{figure*}

\begin{figure*}
\includegraphics[width=0.75\textwidth]{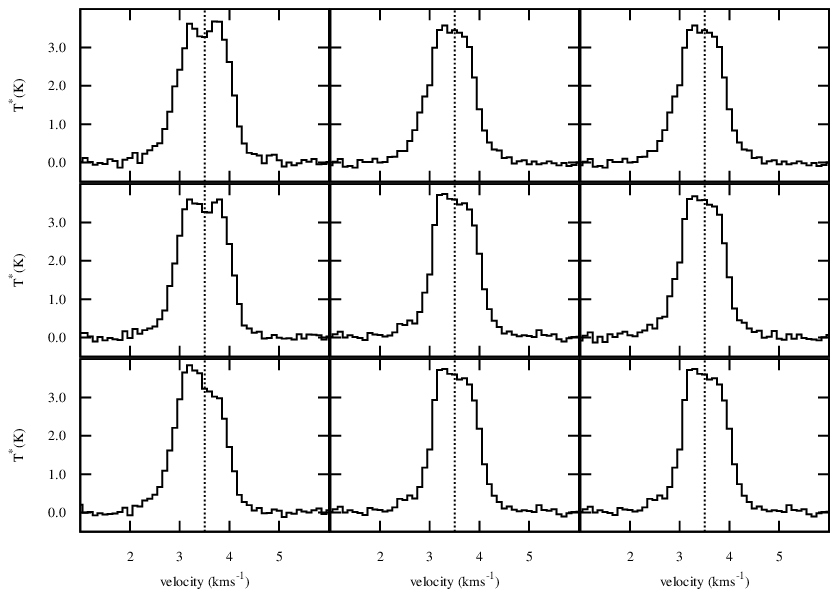}
\caption[Map of HCO$^{+}$ (J=4$\rightarrow$3) spectra in the core A-MM8.]{Map of HCO$^{+}$ (J=4$\rightarrow$3) spectra in the core A-MM8. The central grid of 3$\times$3 HARP pixels is shown here.}
\label{A-MM8}
\end{figure*}

\begin{figure*}
\includegraphics[width=0.75\textwidth]{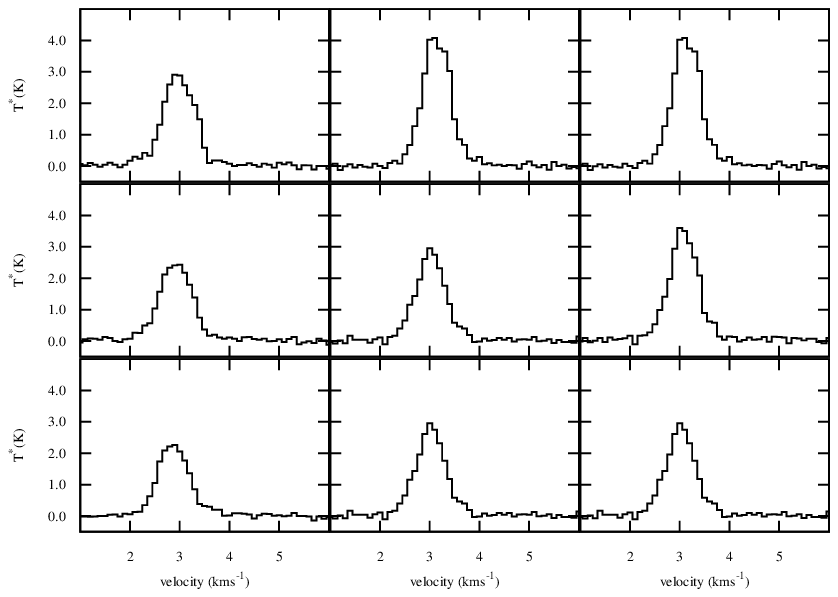}
\caption[Map of HCO$^{+}$ (J=4$\rightarrow$3) spectra in the core A-MM11.]{Map of HCO$^{+}$ (J=4$\rightarrow$3) spectra in the core A-MM11. The central grid of 3$\times$3 HARP pixels is shown here.}
\label{A-MM11}
\end{figure*}

\begin{figure*}
\includegraphics[width=0.75\textwidth]{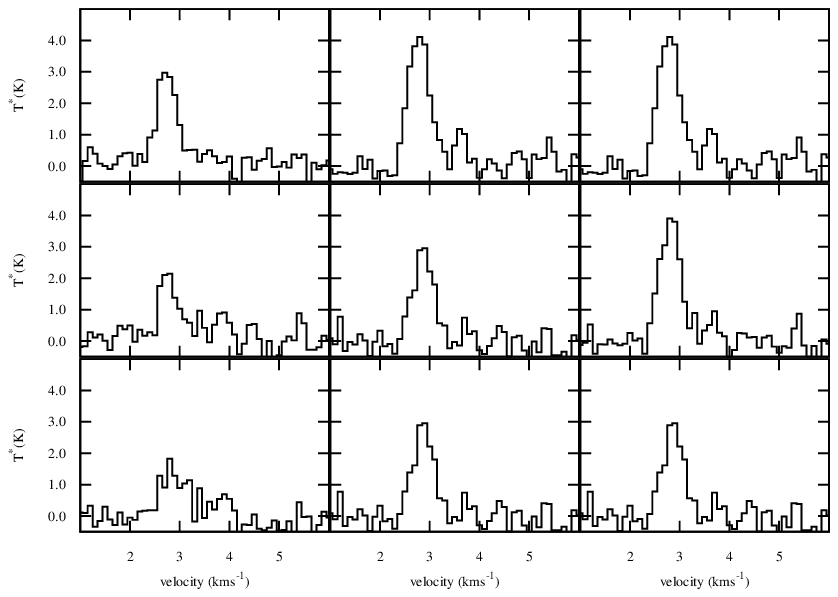}
\caption[Map of HCO$^{+}$ (J=4$\rightarrow$3) spectra in the core A-MM16.]{Map of HCO$^{+}$ (J=4$\rightarrow$3) spectra in the core A-MM16. The central grid of 3$\times$3 HARP pixels is shown here.}
\label{A-MM16}
\end{figure*}

\begin{figure*}
\includegraphics[width=0.75\textwidth]{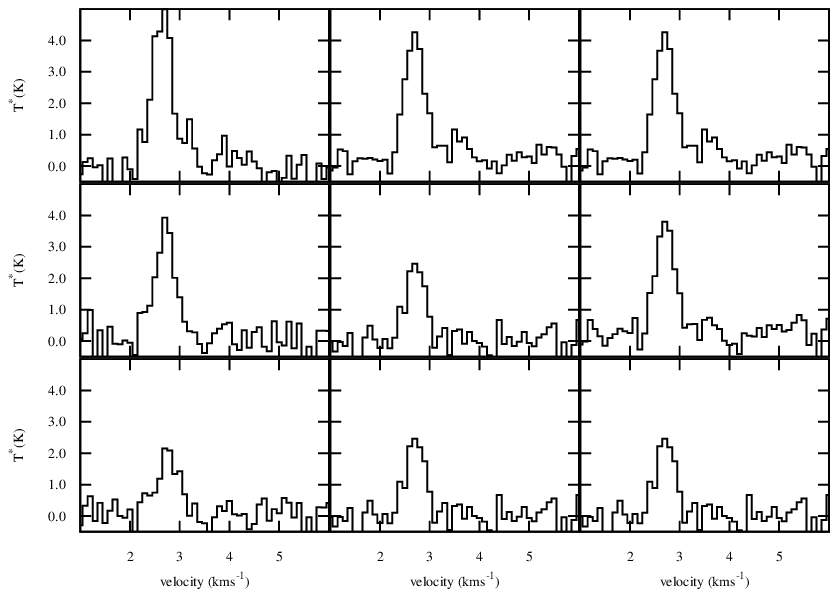}
\caption[Map of HCO$^{+}$ (J=4$\rightarrow$3) spectra in the core A-MM17.]{Map of HCO$^{+}$ (J=4$\rightarrow$3) spectra in the core A-MM17. The central grid of 3$\times$3 HARP pixels is shown here.}
\label{A-MM17}
\end{figure*}

\begin{figure*}
\includegraphics[width=0.75\textwidth]{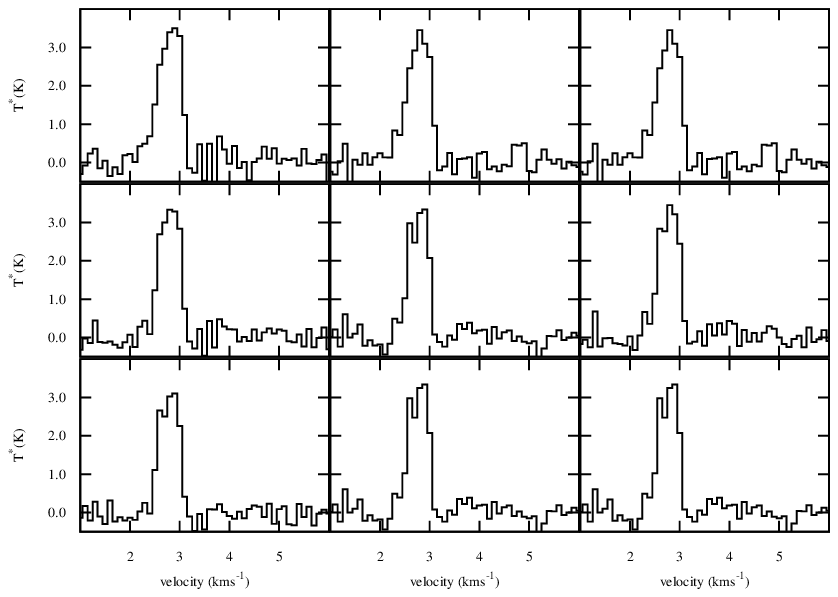}
\caption[Map of HCO$^{+}$ (J=4$\rightarrow$3) spectra in the core A-MM18.]{Map of HCO$^{+}$ (J=4$\rightarrow$3) spectra in the core A-MM18. The central grid of 3$\times$3 HARP pixels is shown here.}
\label{A-MM18}
\end{figure*}

\begin{figure*}
\includegraphics[width=0.75\textwidth]{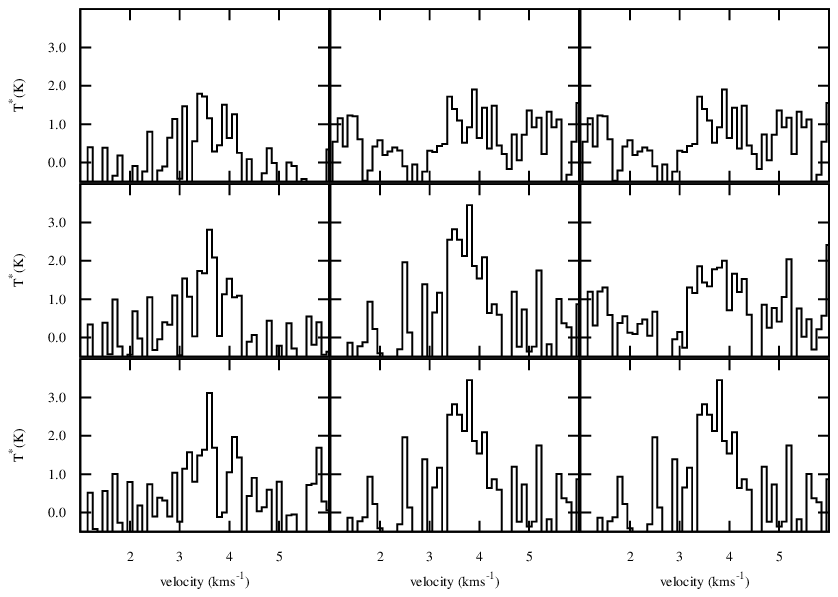}
\caption[Map of HCO$^{+}$ (J=4$\rightarrow$3) spectra in the core A-MM21.]{Map of HCO$^{+}$ (J=4$\rightarrow$3) spectra in the core A-MM21. The central grid of 3$\times$3 HARP pixels is shown here.}
\label{A-MM21}
\end{figure*}

\begin{figure*}
\includegraphics[width=0.75\textwidth]{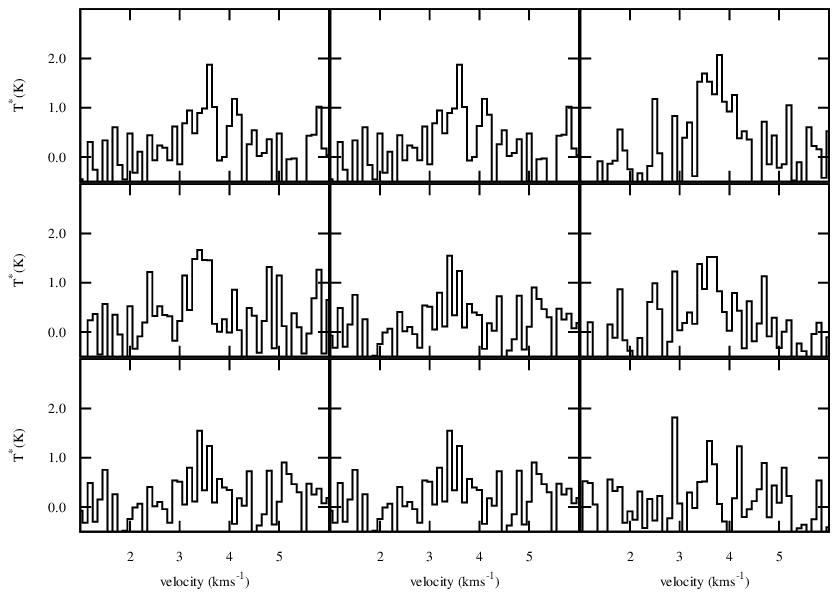}
\caption[Map of HCO$^{+}$ (J=4$\rightarrow$3) spectra in the core A-MM22.]{Map of HCO$^{+}$ (J=4$\rightarrow$3) spectra in the core A-MM22. The central grid of 3$\times$3 HARP pixels is shown here.}
\label{A-MM22}
\end{figure*}

\begin{figure*}
\includegraphics[width=0.75\textwidth]{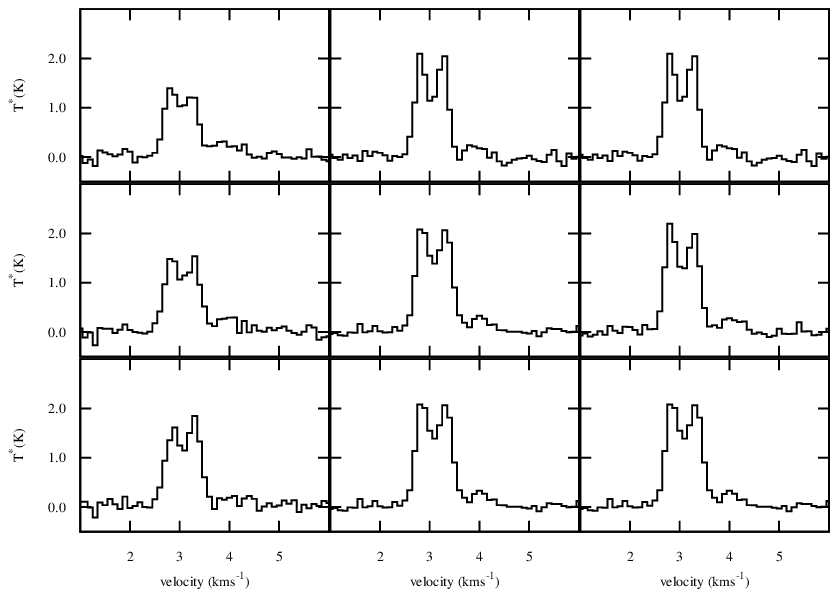}
\caption[Map of HCO$^{+}$ (J=4$\rightarrow$3) spectra in the core A-MM23.]{Map of HCO$^{+}$ (J=4$\rightarrow$3) spectra in the core A-MM23. The central grid of 3$\times$3 HARP pixels is shown here.}
\label{A-MM23}
\end{figure*}

\begin{figure*}
\includegraphics[width=0.75\textwidth]{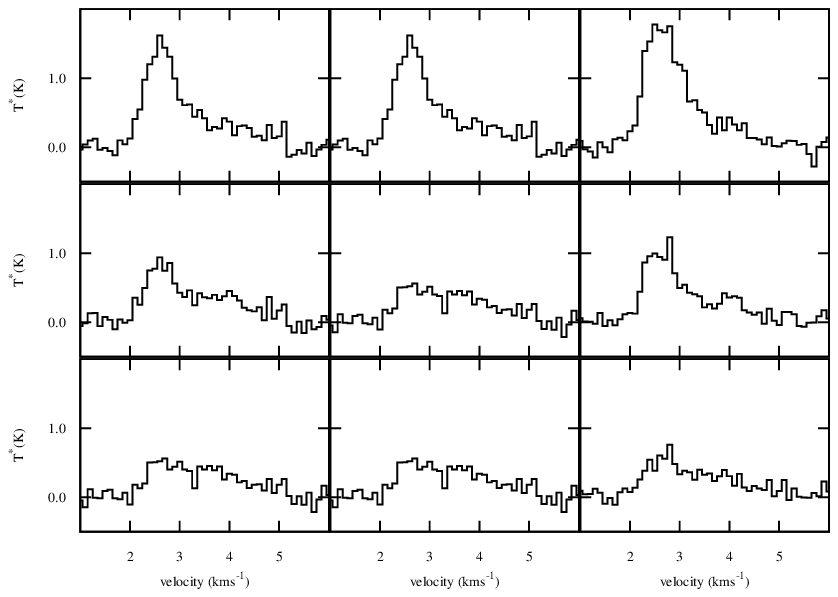}
\caption[Map of HCO$^{+}$ (J=4$\rightarrow$3) spectra in the core A-MM26.]{Map of HCO$^{+}$ (J=4$\rightarrow$3) spectra in the core A-MM26. The central grid of 3$\times$3 HARP pixels is shown here.}
\label{A-MM26}
\end{figure*}

\begin{figure*}
\includegraphics[width=0.75\textwidth]{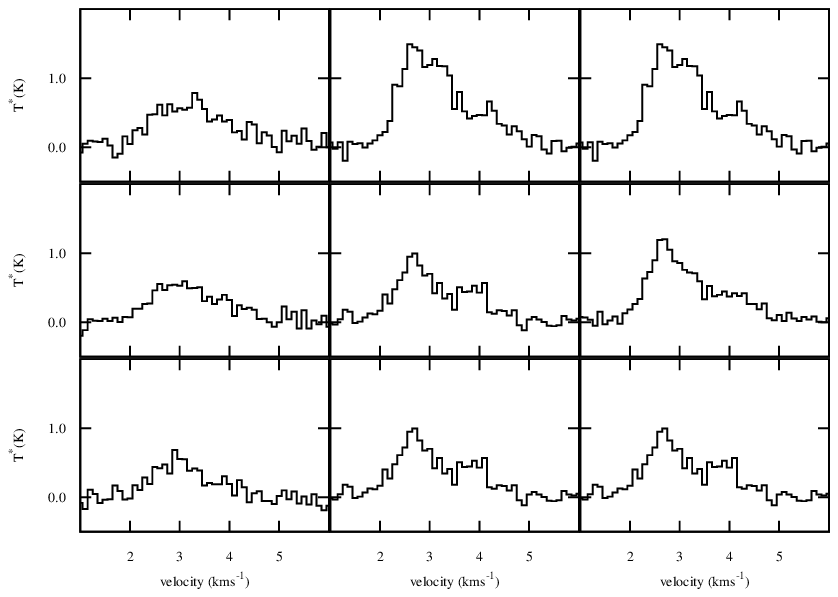}
\caption[Map of HCO$^{+}$ (J=4$\rightarrow$3) spectra in the core A-MM27.]{Map of HCO$^{+}$ (J=4$\rightarrow$3) spectra in the core A-MM27. The central grid of 3$\times$3 HARP pixels is shown here.}
\label{A-MM27}
\end{figure*}

\begin{figure*}
\includegraphics[width=0.75\textwidth]{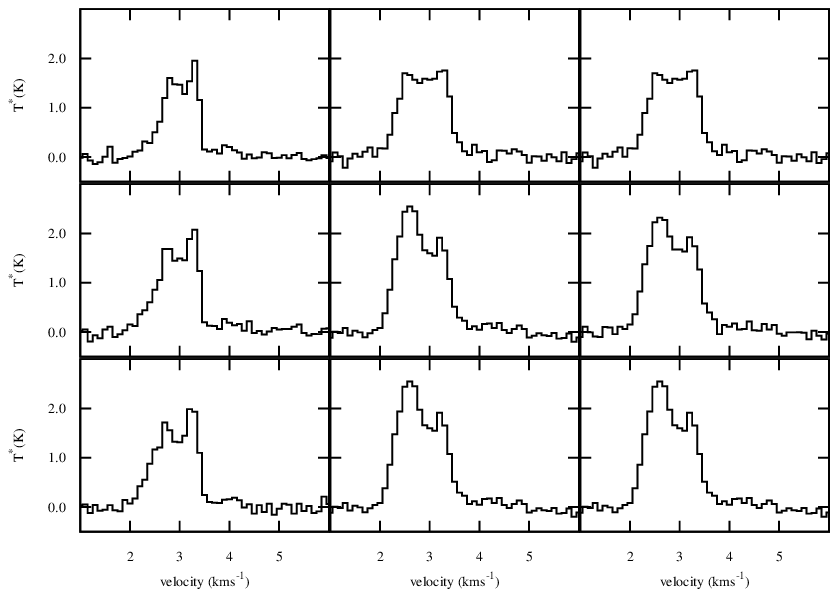}
\caption[Map of HCO$^{+}$ (J=4$\rightarrow$3) spectra in the core A-MM30.]{Map of HCO$^{+}$ (J=4$\rightarrow$3) spectra in the core A-MM30. The central grid of 3$\times$3 HARP pixels is shown here.}
\label{A-MM30}
\end{figure*}

\begin{figure*}
\includegraphics[width=0.75\textwidth]{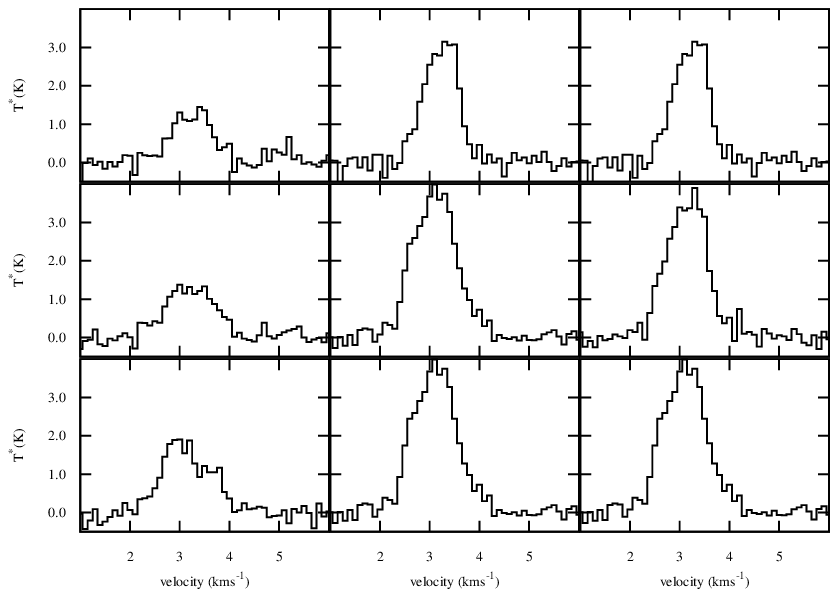}
\caption[Map of HCO$^{+}$ (J=4$\rightarrow$3) spectra in the core A-N.]{Map of HCO$^{+}$ (J=4$\rightarrow$3) spectra in the core A-N. The central grid of 3$\times$3 HARP pixels is shown here.}
\label{A-N}
\end{figure*}

\begin{figure*}
\includegraphics[width=0.75\textwidth]{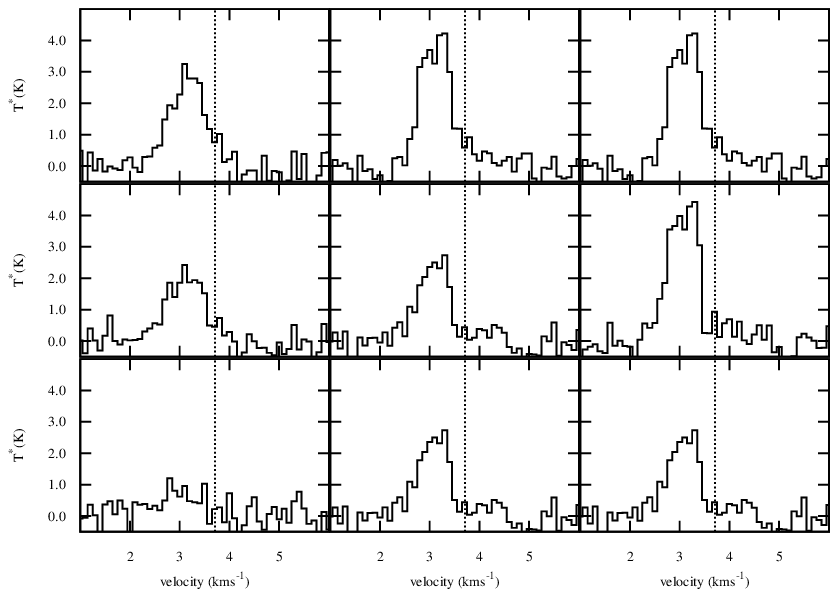}
\caption[Map of HCO$^{+}$ (J=4$\rightarrow$3) spectra in the core A-S.]{Map of HCO$^{+}$ (J=4$\rightarrow$3) spectra in the core A-S. The central grid of 3$\times$3 HARP pixels is shown here.}
\label{A-S}
\end{figure*}

\begin{figure*}
\includegraphics[width=0.75\textwidth]{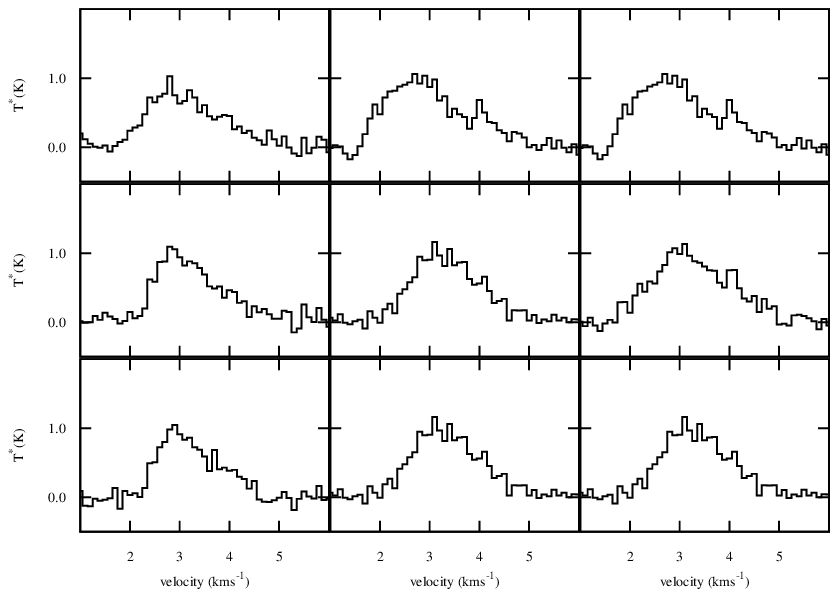}
\caption[Map of HCO$^{+}$ (J=4$\rightarrow$3) spectra in the core A2-MM1.]{Map of HCO$^{+}$ (J=4$\rightarrow$3) spectra in the core A2-MM1. The central grid of 3$\times$3 HARP pixels is shown here.}
\label{A2-MM1}
\end{figure*}

\begin{figure*}
\includegraphics[width=0.75\textwidth]{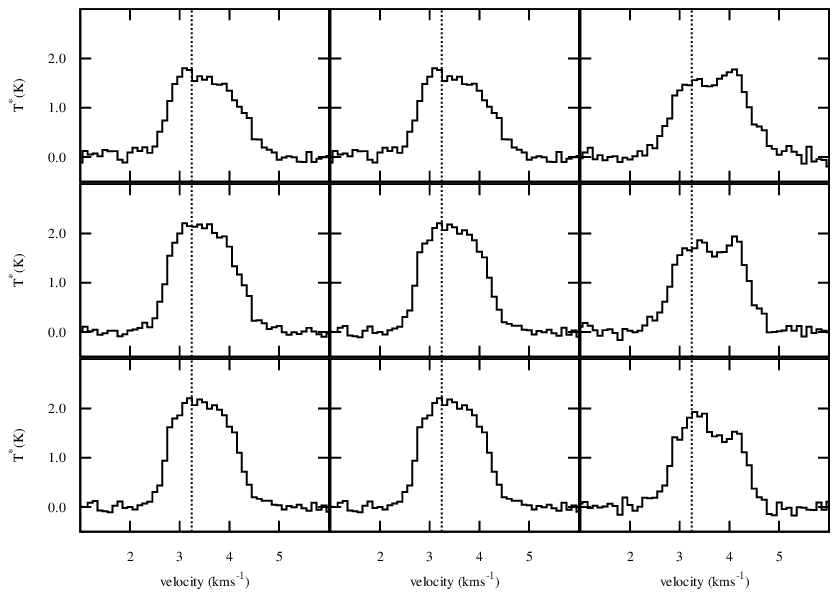}
\caption[Map of HCO$^{+}$ (J=4$\rightarrow$3) spectra in the core A3-MM1.]{Map of HCO$^{+}$ (J=4$\rightarrow$3) spectra in the core A3-MM1. The central grid of 3$\times$3 HARP pixels is shown here.}
\label{A3-MM1}
\end{figure*}

\begin{figure*}
\includegraphics[width=0.75\textwidth]{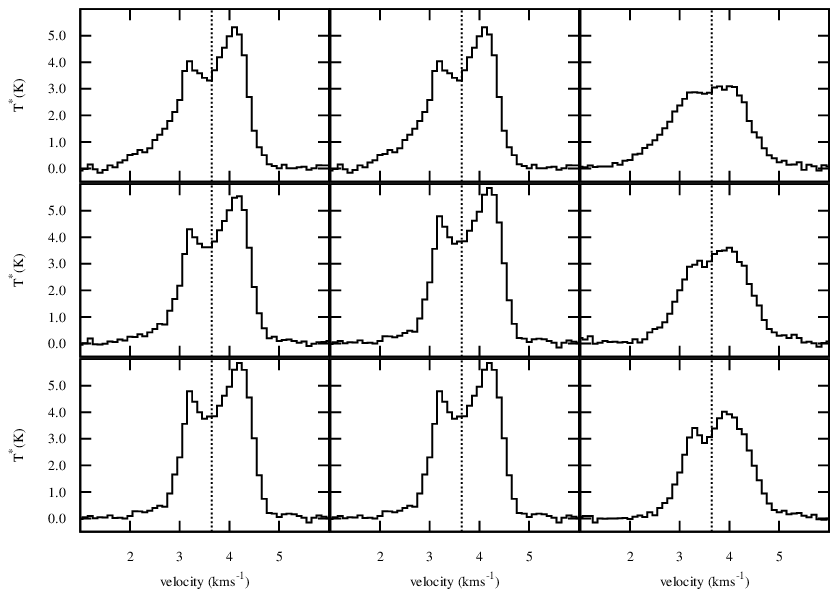}
\caption[Map of HCO$^{+}$ (J=4$\rightarrow$3) spectra in the core SM1.]{Map of HCO$^{+}$ (J=4$\rightarrow$3) spectra in the core SM1. The central grid of 3$\times$3 HARP pixels is shown here.}
\label{SM1}
\end{figure*}

\begin{figure*}
\includegraphics[width=0.75\textwidth]{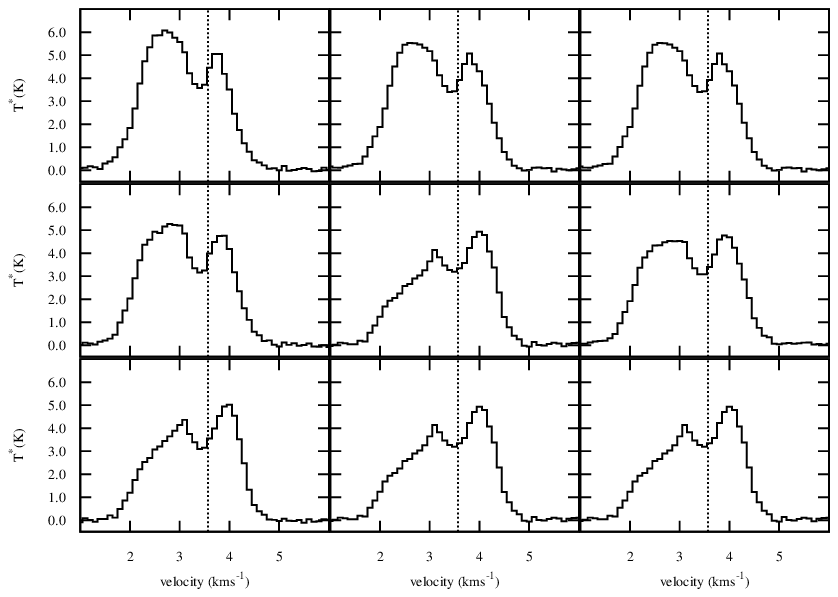}
\caption[Map of HCO$^{+}$ (J=4$\rightarrow$3) spectra in the core SM1N.]{Map of HCO$^{+}$ (J=4$\rightarrow$3) spectra in the core SM1N. The central grid of 3$\times$3 HARP pixels is shown here.}
\label{SM1N}
\end{figure*}

\begin{figure*}
\includegraphics[width=0.75\textwidth]{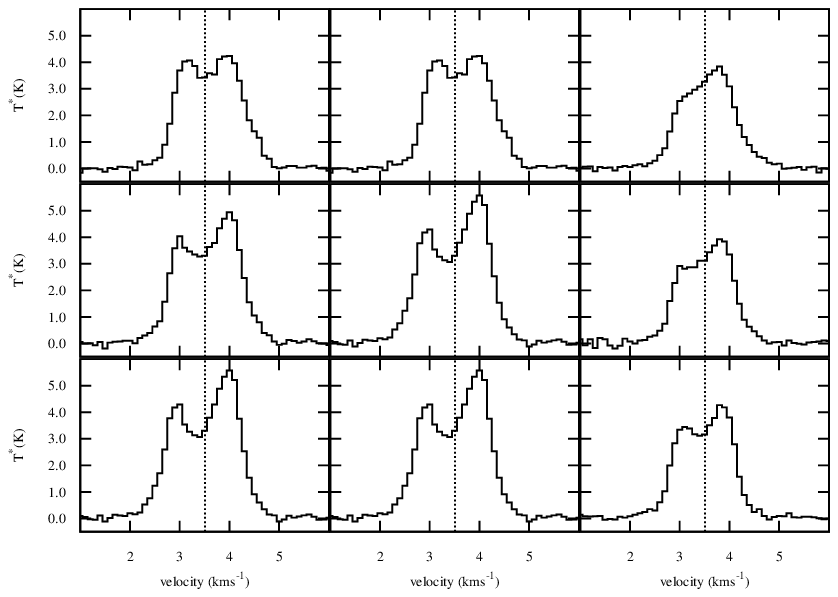}
\caption[Map of HCO$^{+}$ (J=4$\rightarrow$3) spectra in the core SM2.]{Map of HCO$^{+}$ (J=4$\rightarrow$3) spectra in the core SM2. The central grid of 3$\times$3 HARP pixels is shown here.}
\label{SM2}
\end{figure*}

\begin{figure*}
\includegraphics[width=0.75\textwidth]{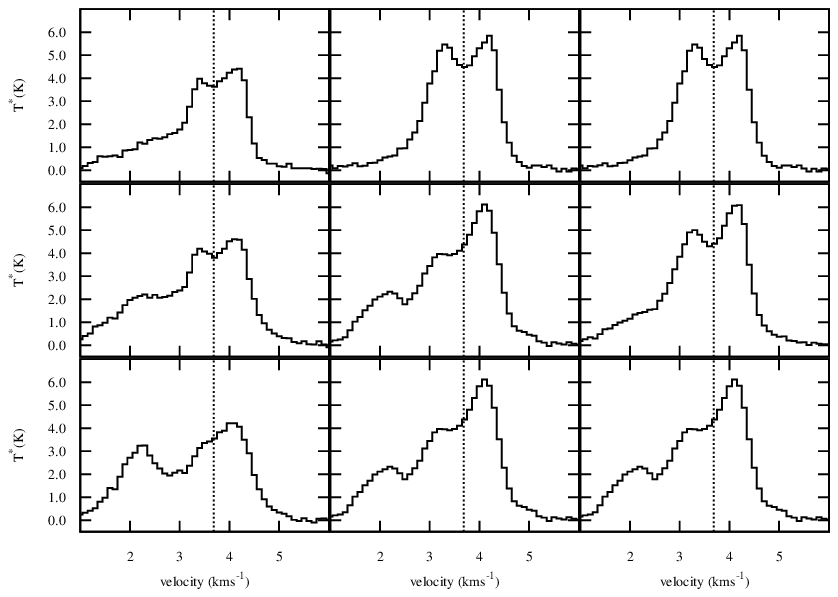}
\caption[Map of HCO$^{+}$ (J=4$\rightarrow$3) spectra in the core VLA1623.]{Map of HCO$^{+}$ (J=4$\rightarrow$3) spectra in the core VLA1623. The central grid of 3$\times$3 HARP pixels is shown here.}
\label{VLA1623}
\end{figure*}

\begin{figure*}
\includegraphics[width=0.75\textwidth]{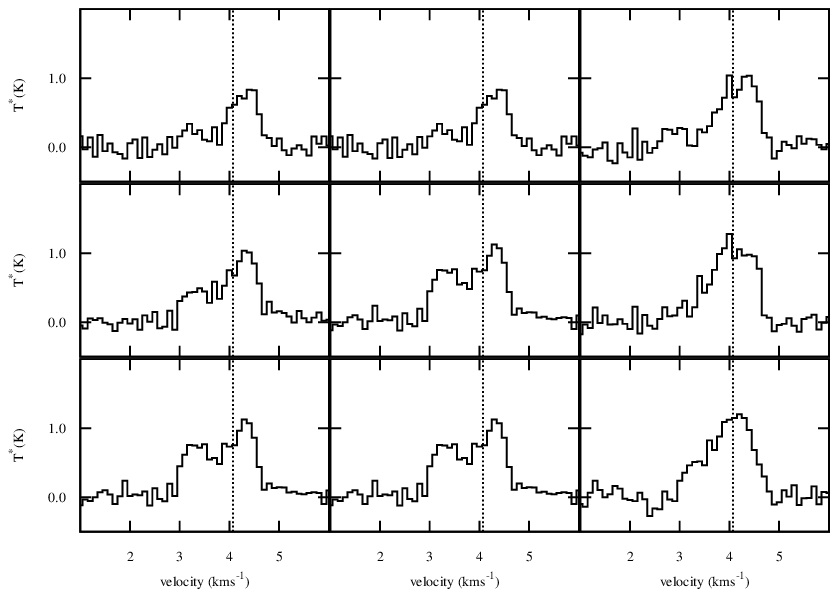}
\caption[Map of HCO$^{+}$ (J=4$\rightarrow$3) spectra in the core B1-MM1.]{Map of HCO$^{+}$ (J=4$\rightarrow$3) spectra in the core B1-MM1. The central grid of 3$\times$3 HARP pixels is shown here.}
\label{B1-MM1}
\end{figure*}

\begin{figure*}
\includegraphics[width=0.75\textwidth]{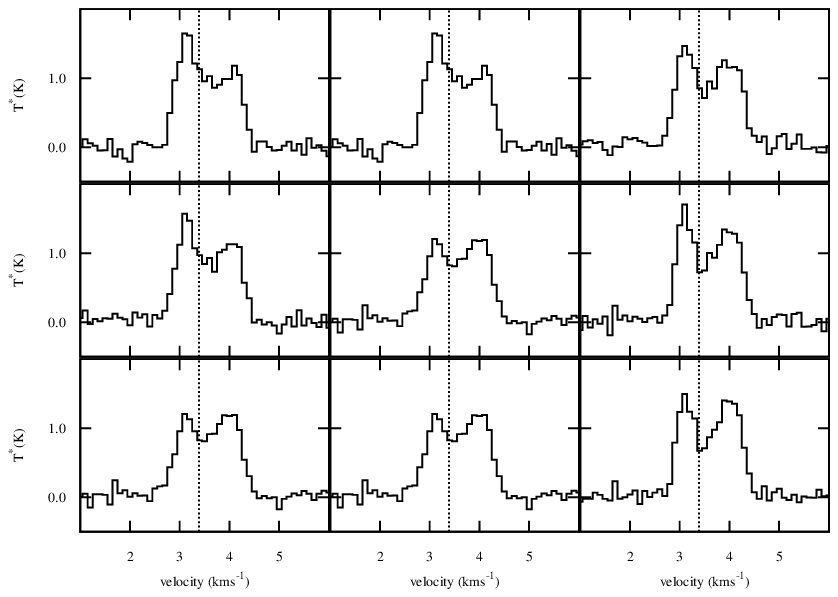}
\caption[Map of HCO$^{+}$ (J=4$\rightarrow$3) spectra in the core B1-MM2.]{Map of HCO$^{+}$ (J=4$\rightarrow$3) spectra in the core B1-MM2. The central grid of 3$\times$3 HARP pixels is shown here.}
\label{B1-MM2}
\end{figure*}

\begin{figure*}
\includegraphics[width=0.75\textwidth]{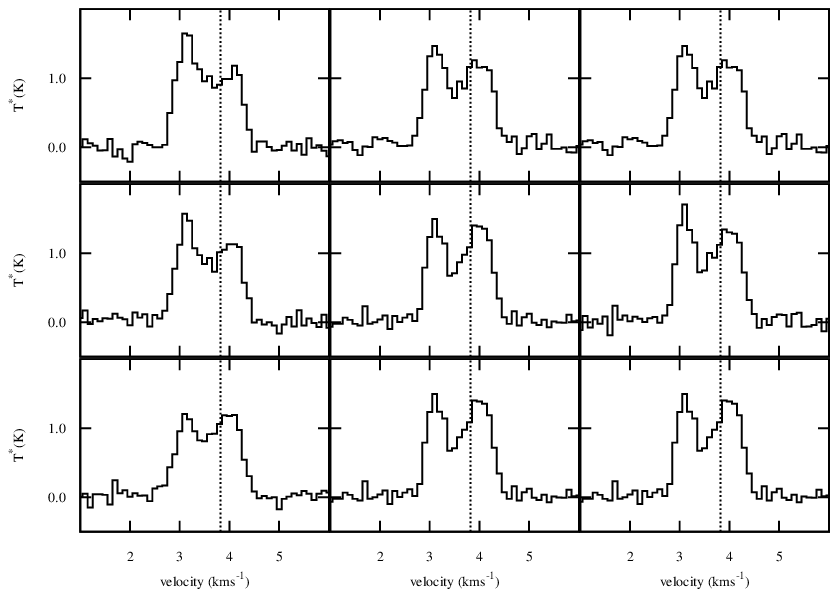}
\caption[Map of HCO$^{+}$ (J=4$\rightarrow$3) spectra in the core B1-MM3.]{Map of HCO$^{+}$ (J=4$\rightarrow$3) spectra in the core B1-MM3. The central grid of 3$\times$3 HARP pixels is shown here.}
\label{B1-MM3}
\end{figure*}

\begin{figure*}
\includegraphics[width=0.75\textwidth]{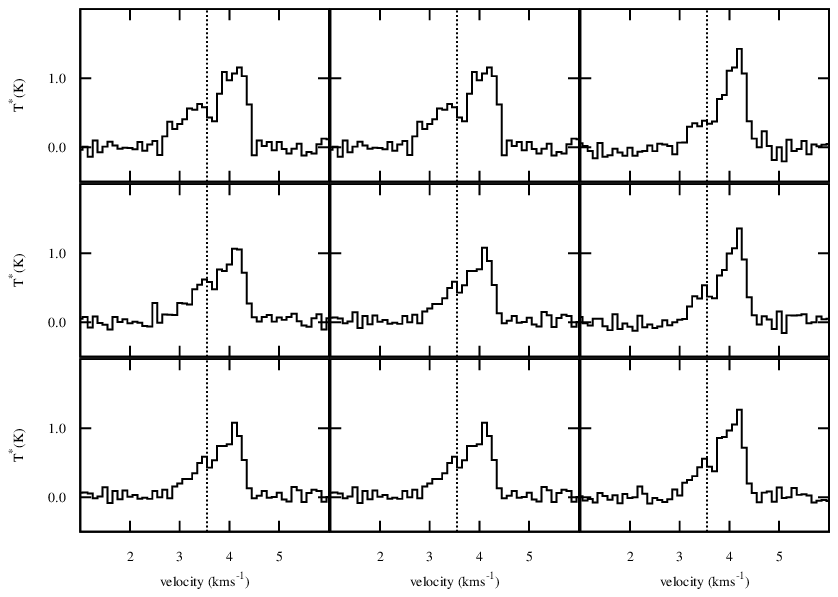}
\caption[Map of HCO$^{+}$ (J=4$\rightarrow$3) spectra in the core B1-MM4.]{Map of HCO$^{+}$ (J=4$\rightarrow$3) spectra in the core B1-MM4. The central grid of 3$\times$3 HARP pixels is shown here.}
\label{B1-MM4}
\end{figure*}

\begin{figure*}
\includegraphics[width=0.75\textwidth]{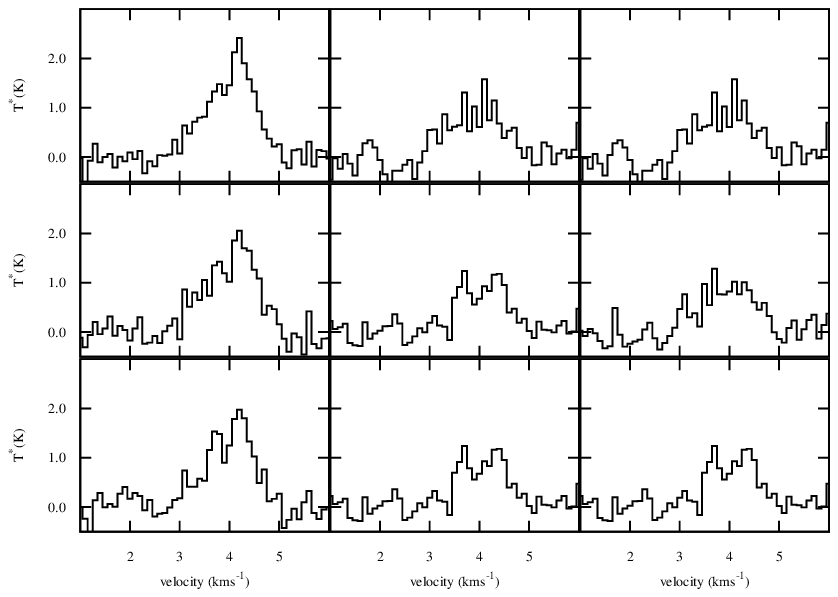}
\caption[Map of HCO$^{+}$ (J=4$\rightarrow$3) spectra in the core B1-MM7.]{Map of HCO$^{+}$ (J=4$\rightarrow$3) spectra in the core B1-MM7. The central grid of 3$\times$3 HARP pixels is shown here.}
\label{B1-MM7}
\end{figure*}

\begin{figure*}
\includegraphics[width=0.75\textwidth]{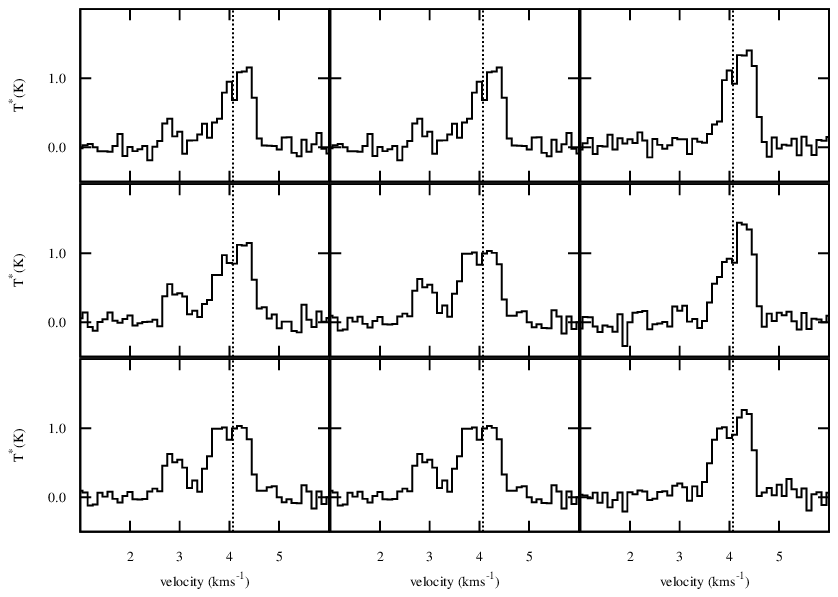}
\caption[Map of HCO$^{+}$ (J=4$\rightarrow$3) spectra in the core B1B2-MM1.]{Map of HCO$^{+}$ (J=4$\rightarrow$3) spectra in the core B1B2-MM1. The central grid of 3$\times$3 HARP pixels is shown here.}
\label{B1B2-MM1}
\end{figure*}

\begin{figure*}
\includegraphics[width=0.75\textwidth]{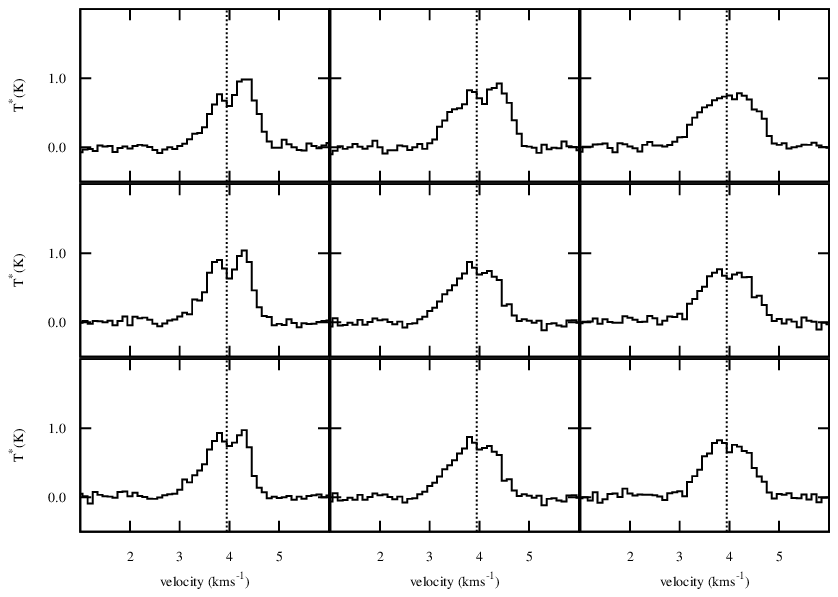}
\caption[Map of HCO$^{+}$ (J=4$\rightarrow$3) spectra in the core B2-MM2.]{Map of HCO$^{+}$ (J=4$\rightarrow$3) spectra in the core B2-MM2. The central grid of 3$\times$3 HARP pixels is shown here.}
\label{B2-MM2}
\end{figure*}

\begin{figure*}
\includegraphics[width=0.75\textwidth]{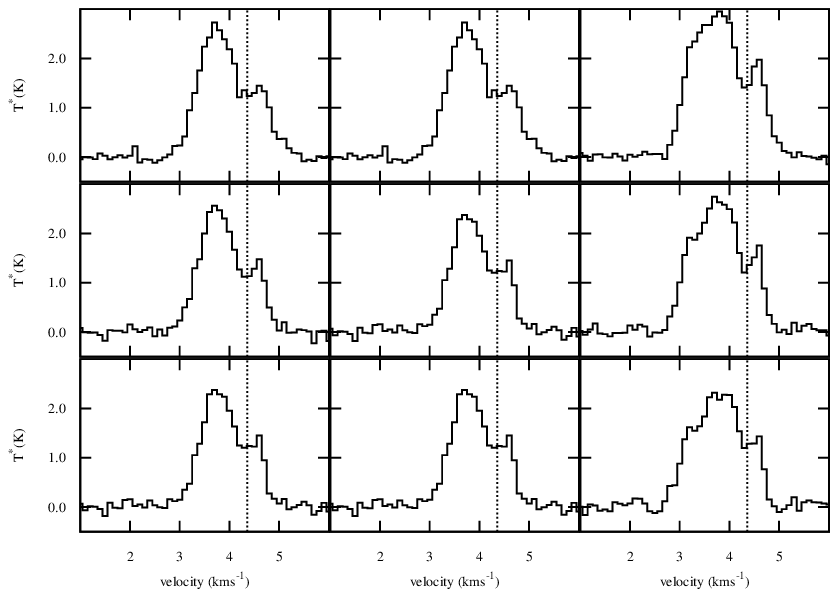}
\caption[Map of HCO$^{+}$ (J=4$\rightarrow$3) spectra in the core B2-MM4.]{Map of HCO$^{+}$ (J=4$\rightarrow$3) spectra in the core B2-MM4. The central grid of 3$\times$3 HARP pixels is shown here.}
\label{B2-MM4}
\end{figure*}

\begin{figure*}
\includegraphics[width=0.75\textwidth]{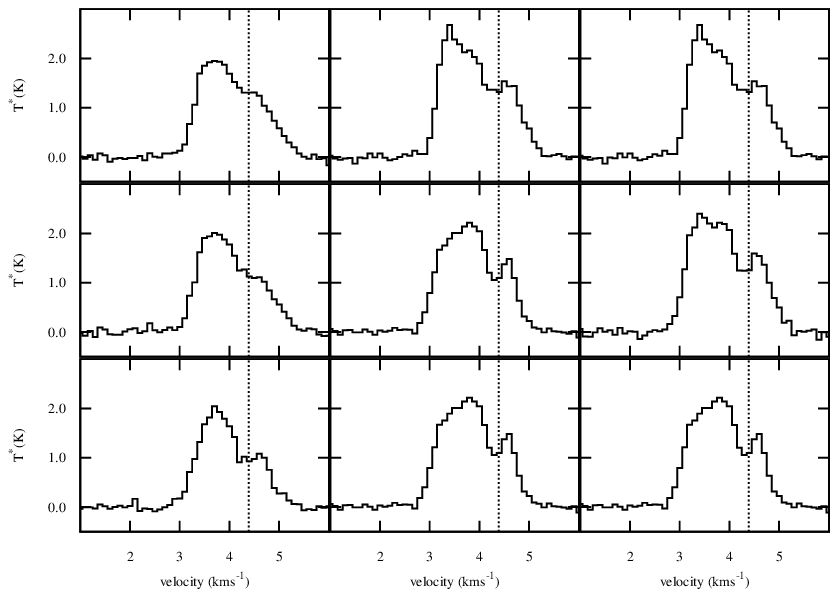}
\caption[Map of HCO$^{+}$ (J=4$\rightarrow$3) spectra in the core B2-MM5.]{Map of HCO$^{+}$ (J=4$\rightarrow$3) spectra in the core B2-MM5. The central grid of 3$\times$3 HARP pixels is shown here.}
\label{B2-MM5}
\end{figure*}

\begin{figure*}
\includegraphics[width=0.75\textwidth]{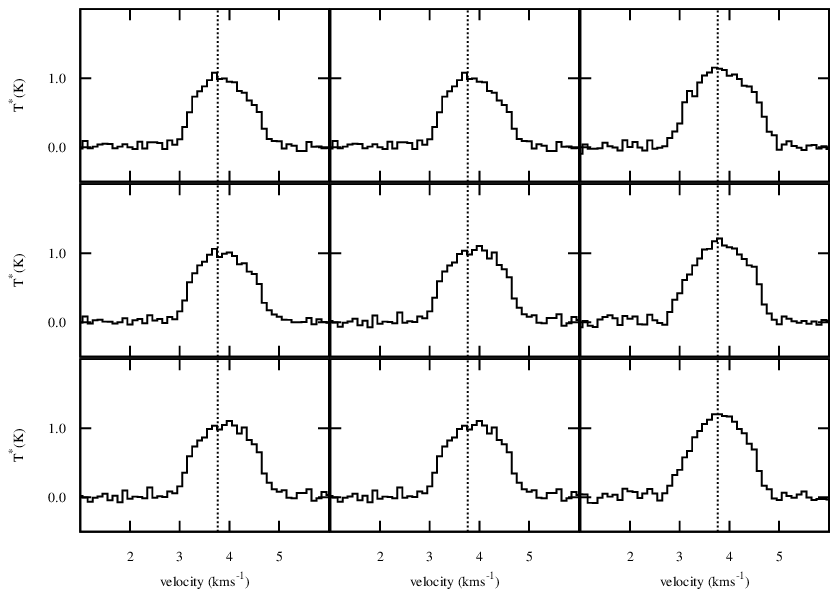}
\caption[Map of HCO$^{+}$ (J=4$\rightarrow$3) spectra in the core B2-MM6.]{Map of HCO$^{+}$ (J=4$\rightarrow$3) spectra in the core B2-MM6. The central grid of 3$\times$3 HARP pixels is shown here.}
\label{B2-MM6}
\end{figure*}

\begin{figure*}
\includegraphics[width=0.75\textwidth]{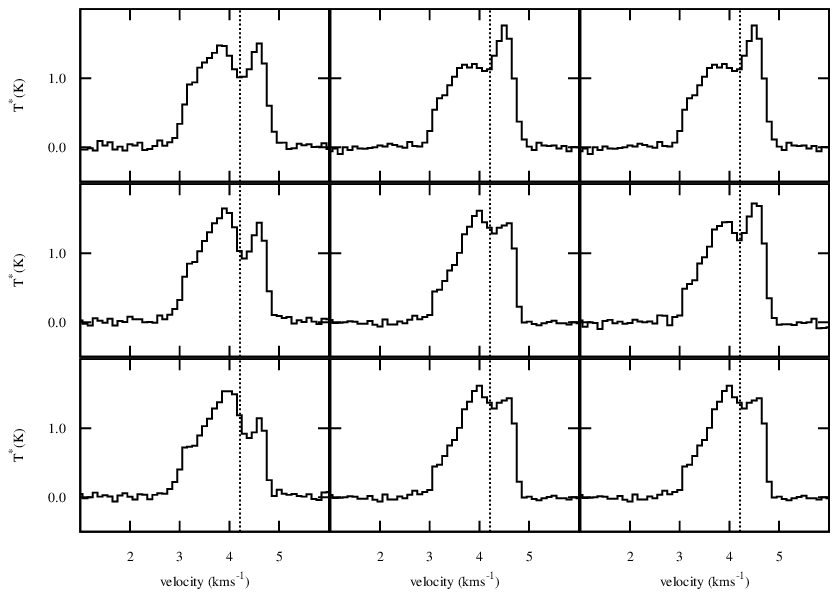}
\caption[Map of HCO$^{+}$ (J=4$\rightarrow$3) spectra in the core B2-MM7.]{Map of HCO$^{+}$ (J=4$\rightarrow$3) spectra in the core B2-MM7. The central grid of 3$\times$3 HARP pixels is shown here.}
\label{B2-MM7}
\end{figure*}

\begin{figure*}
\includegraphics[width=0.75\textwidth]{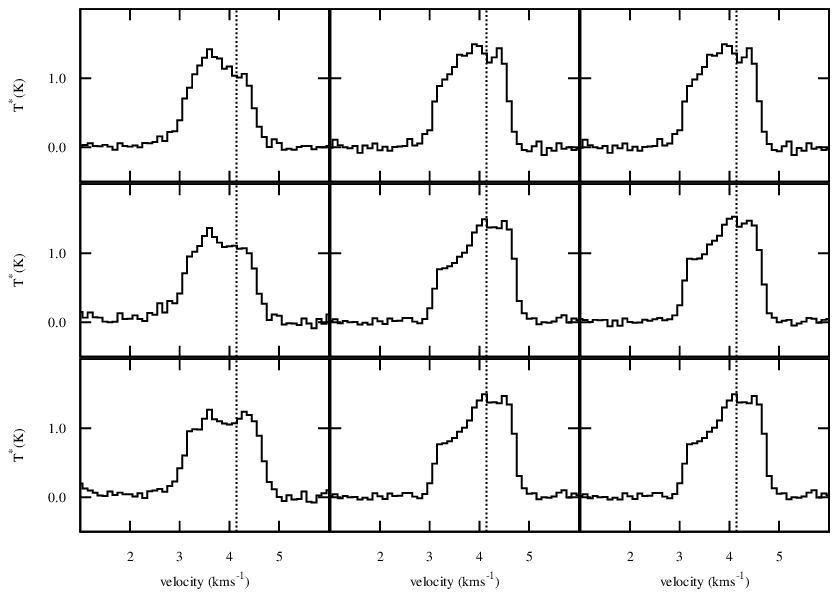}
\caption[Map of HCO$^{+}$ (J=4$\rightarrow$3) spectra in the core B2-MM8.]{Map of HCO$^{+}$ (J=4$\rightarrow$3) spectra in the core B2-MM8. The central grid of 3$\times$3 HARP pixels is shown here.}
\label{B2-MM8}
\end{figure*}

\begin{figure*}
\includegraphics[width=0.75\textwidth]{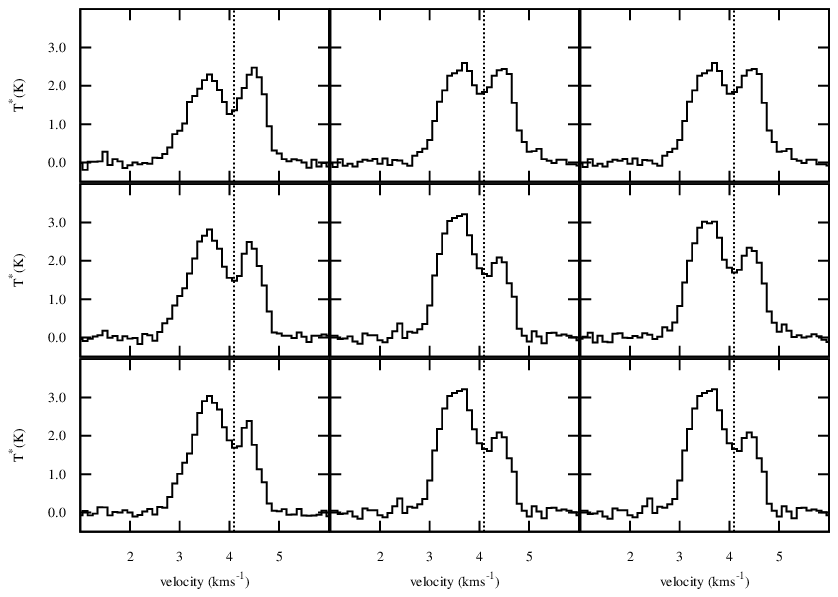}
\caption[Map of HCO$^{+}$ (J=4$\rightarrow$3) spectra in the core B2-MM9.]{Map of HCO$^{+}$ (J=4$\rightarrow$3) spectra in the core B2-MM9. The central grid of 3$\times$3 HARP pixels is shown here.}
\label{B2-MM9}
\end{figure*}

\begin{figure*}
\includegraphics[width=0.75\textwidth]{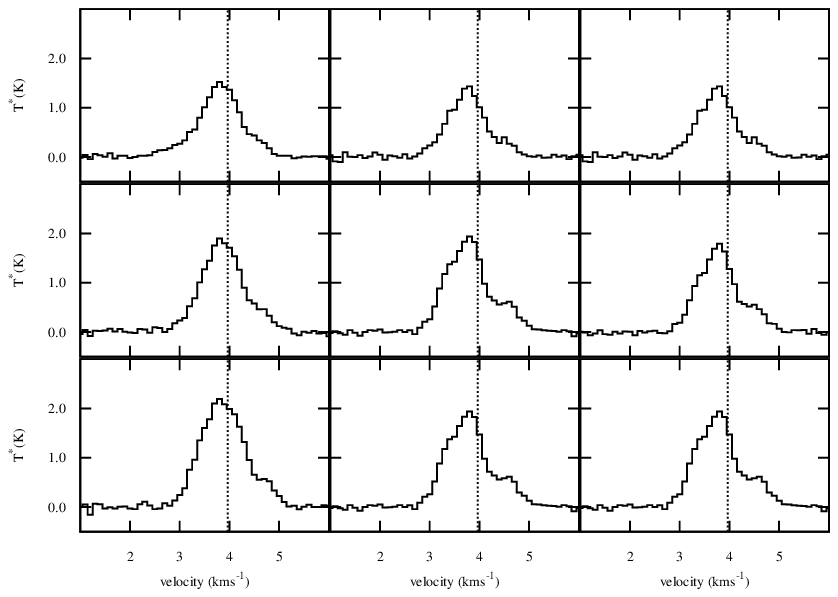}
\caption[Map of HCO$^{+}$ (J=4$\rightarrow$3) spectra in the core B2-MM13.]{Map of HCO$^{+}$ (J=4$\rightarrow$3) spectra in the core B2-MM13. The central grid of 3$\times$3 HARP pixels is shown here.}
\label{B2-MM13}
\end{figure*}

\begin{figure*}
\includegraphics[width=0.75\textwidth]{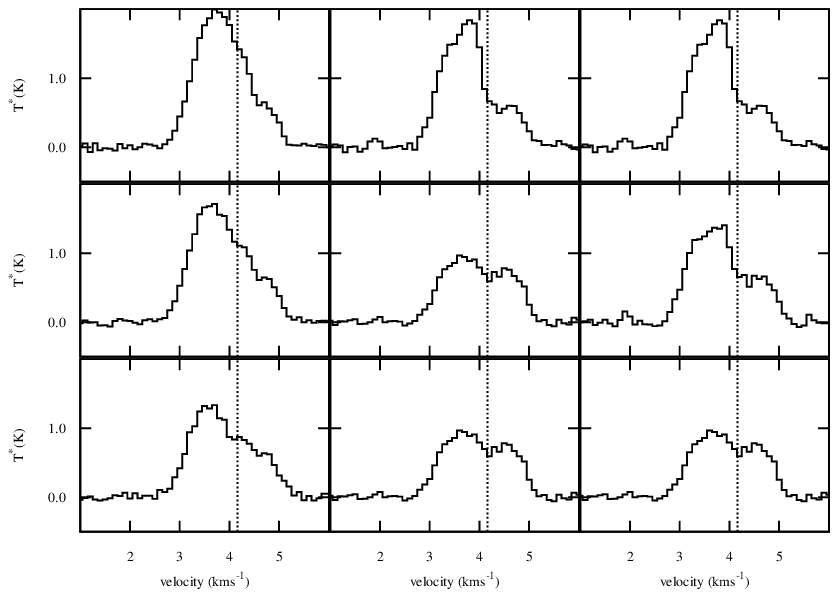}
\caption[Map of HCO$^{+}$ (J=4$\rightarrow$3) spectra in the core B2-MM14.]{Map of HCO$^{+}$ (J=4$\rightarrow$3) spectra in the core B2-MM14. The central grid of 3$\times$3 HARP pixels is shown here.}
\label{B2-MM14}
\end{figure*}

\begin{figure*}
\includegraphics[width=0.75\textwidth]{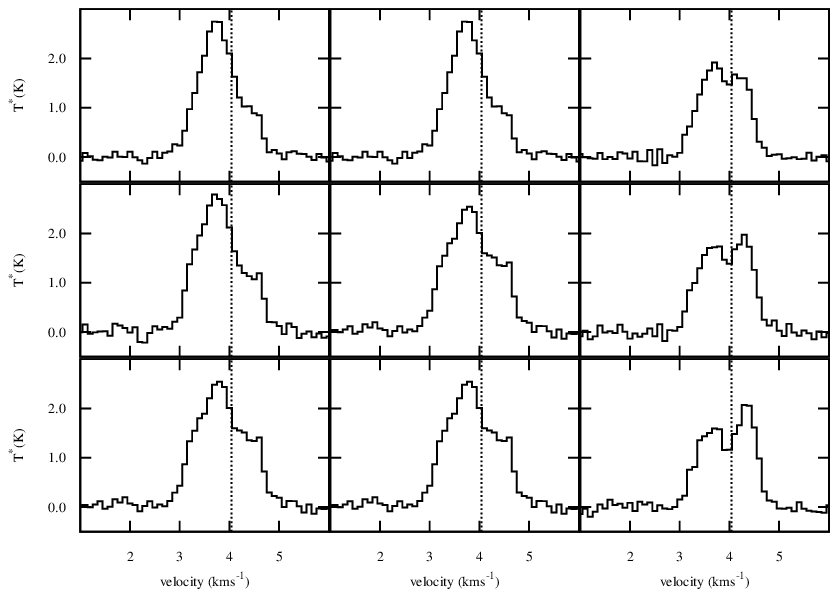}
\caption[Map of HCO$^{+}$ (J=4$\rightarrow$3) spectra in the core B2-MM16.]{Map of HCO$^{+}$ (J=4$\rightarrow$3) spectra in the core B2-MM16. The central grid of 3$\times$3 HARP pixels is shown here.}
\label{B2-MM16}
\end{figure*}

\begin{figure*}
\includegraphics[width=0.75\textwidth]{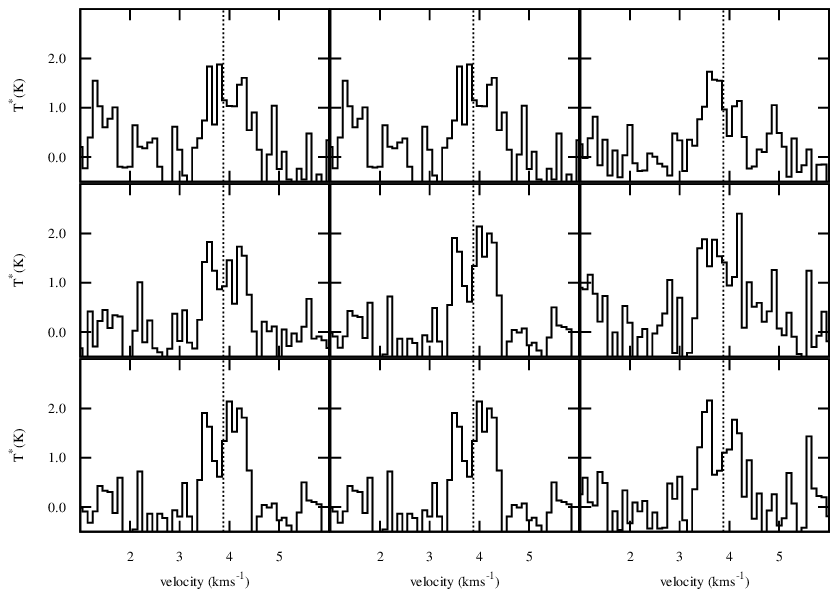}
\caption[Map of HCO$^{+}$ (J=4$\rightarrow$3) spectra in the core C-MM2.]{Map of HCO$^{+}$ (J=4$\rightarrow$3) spectra in the core C-MM2. The central grid of 3$\times$3 HARP pixels is shown here.}
\label{C-MM2}
\end{figure*}

\begin{figure*}
\includegraphics[width=0.75\textwidth]{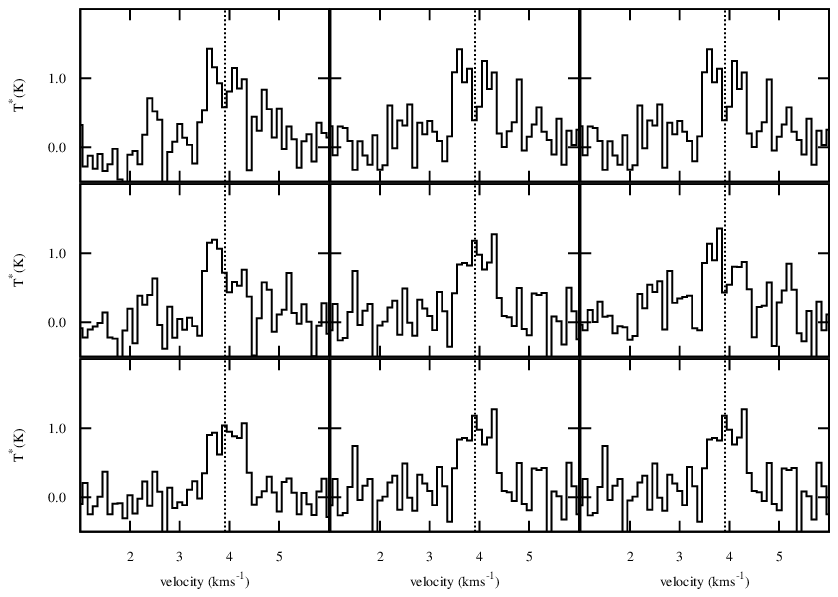}
\caption[Map of HCO$^{+}$ (J=4$\rightarrow$3) spectra in the core C-MM3.]{Map of HCO$^{+}$ (J=4$\rightarrow$3) spectra in the core C-MM3. The central grid of 3$\times$3 HARP pixels is shown here.}
\label{C-MM3}
\end{figure*}

\begin{figure*}
\includegraphics[width=0.75\textwidth]{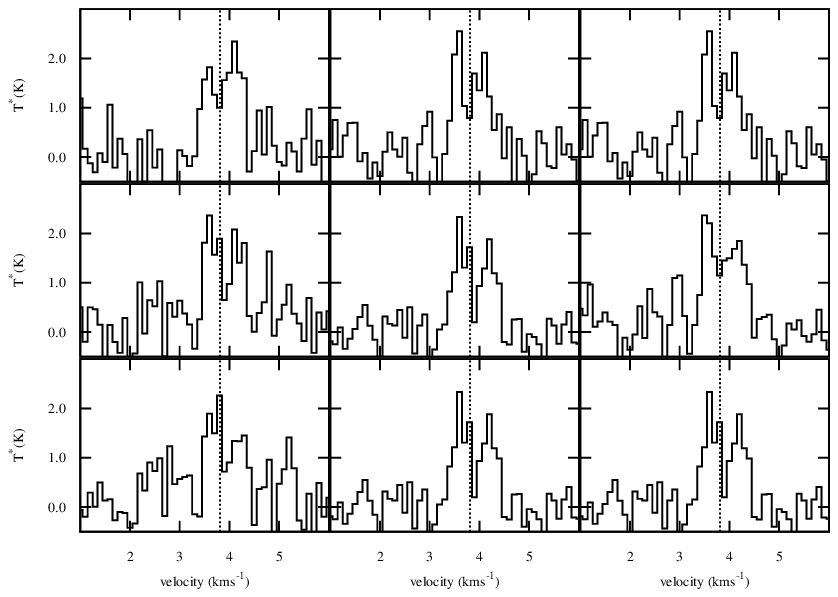}
\caption[Map of HCO$^{+}$ (J=4$\rightarrow$3) spectra in the core C-MM5.]{Map of HCO$^{+}$ (J=4$\rightarrow$3) spectra in the core C-MM5. The central grid of 3$\times$3 HARP pixels is shown here.}
\label{C-MM5}
\end{figure*}

\begin{figure*}
\includegraphics[width=0.75\textwidth]{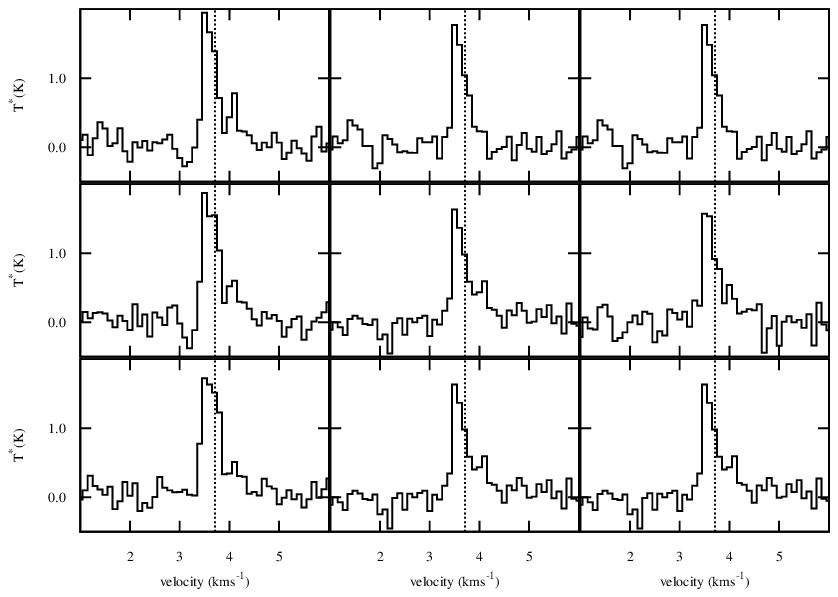}
\caption[Map of HCO$^{+}$ (J=4$\rightarrow$3) spectra in the core C-MM6.]{Map of HCO$^{+}$ (J=4$\rightarrow$3) spectra in the core C-MM6. The central grid of 3$\times$3 HARP pixels is shown here.}
\label{C-MM6}
\end{figure*}

\begin{figure*}
\includegraphics[width=0.75\textwidth]{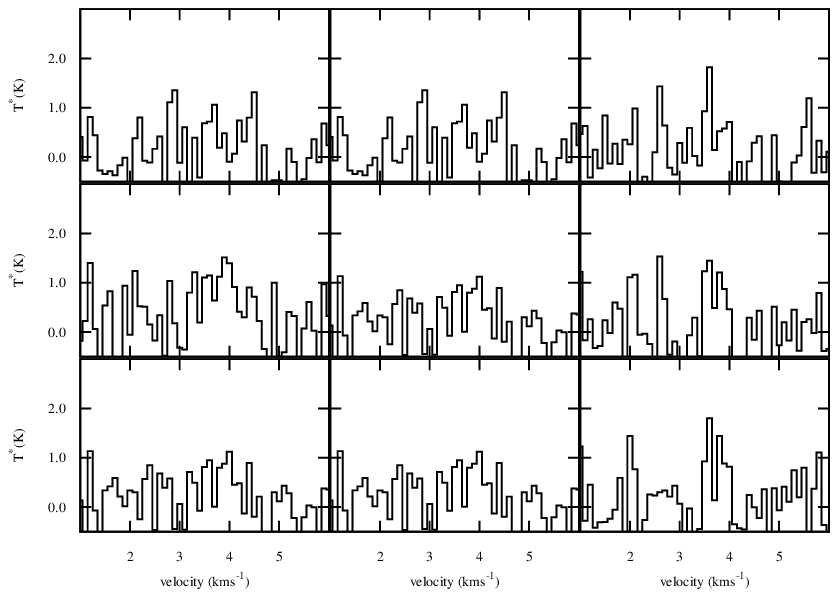}
\caption[Map of HCO$^{+}$ (J=4$\rightarrow$3) spectra in the core C-MM8.]{Map of HCO$^{+}$ (J=4$\rightarrow$3) spectra in the core C-MM8. The central grid of 3$\times$3 HARP pixels is shown here.}
\label{C-MM8}
\end{figure*}

\begin{figure*}
\includegraphics[width=0.75\textwidth]{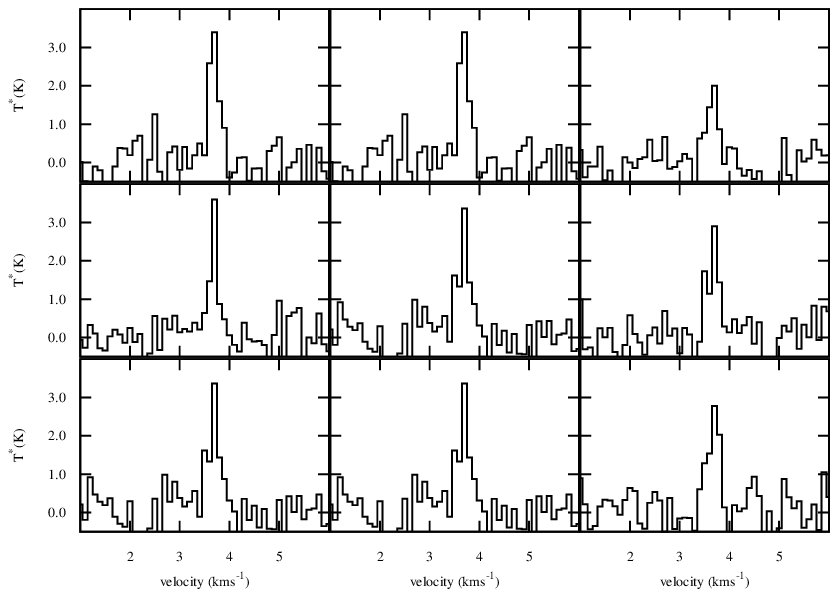}
\caption[Map of HCO$^{+}$ (J=4$\rightarrow$3) spectra in the core C-MM9.]{Map of HCO$^{+}$ (J=4$\rightarrow$3) spectra in the core C-MM9. The central grid of 3$\times$3 HARP pixels is shown here.}
\label{C-MM9}
\end{figure*}

\begin{figure*}
\includegraphics[width=0.75\textwidth]{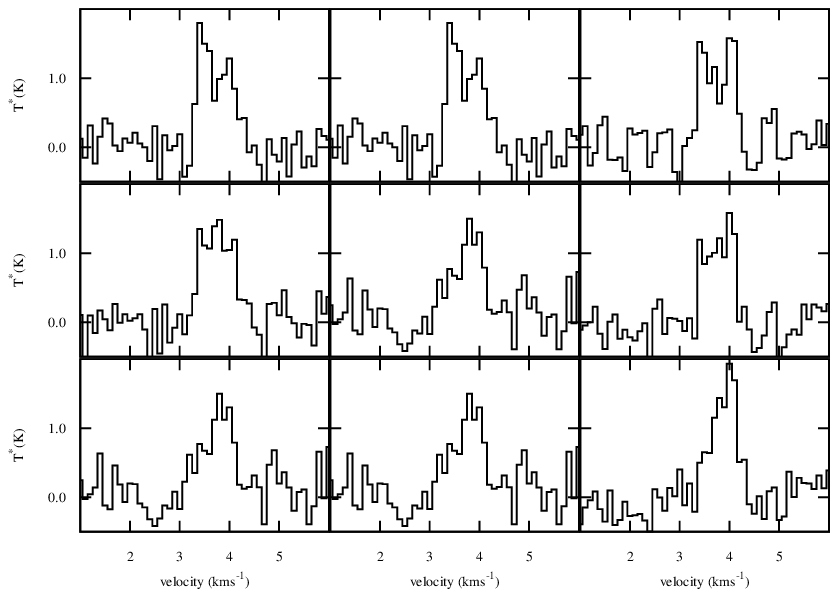}
\caption[Map of HCO$^{+}$ (J=4$\rightarrow$3) spectra in the core C-MM10.]{Map of HCO$^{+}$ (J=4$\rightarrow$3) spectra in the core C-MM10. The central grid of 3$\times$3 HARP pixels is shown here.}
\label{C-MM10}
\end{figure*}

\begin{figure*}
\includegraphics[width=0.75\textwidth]{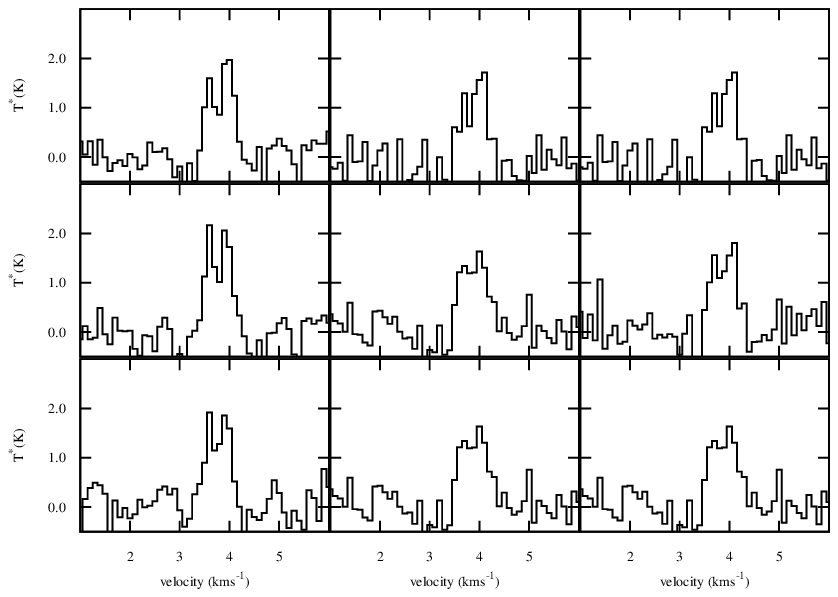}
\caption[Map of HCO$^{+}$ (J=4$\rightarrow$3) spectra in the core C-MM12.]{Map of HCO$^{+}$ (J=4$\rightarrow$3) spectra in the core C-MM12. The central grid of 3$\times$3 HARP pixels is shown here.}
\label{C-MM12}
\end{figure*}

\begin{figure*}
\includegraphics[width=0.75\textwidth]{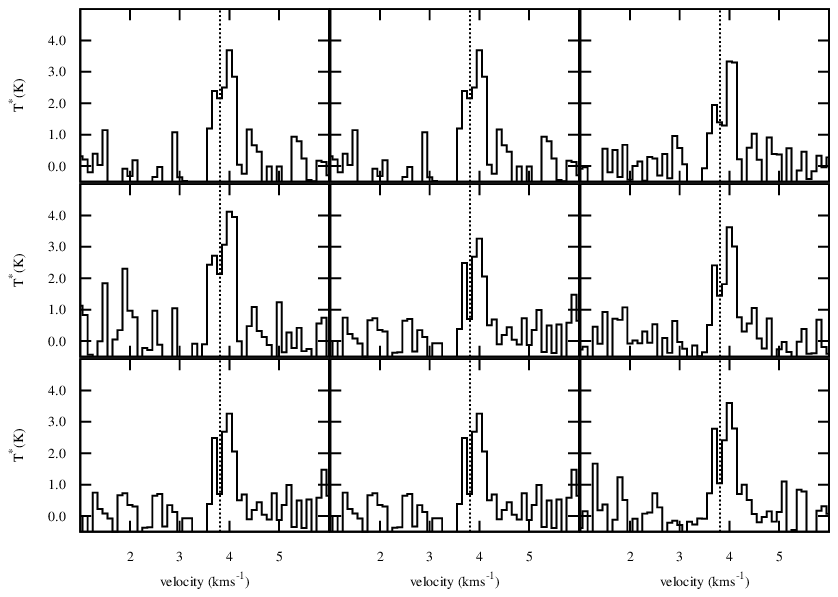}
\caption[Map of HCO$^{+}$ (J=4$\rightarrow$3) spectra in the core C-N.]{Map of HCO$^{+}$ (J=4$\rightarrow$3) spectra in the core C-N. The central grid of 3$\times$3 HARP pixels is shown here.}
\label{C-N}
\end{figure*}

\begin{figure*}
\includegraphics[width=0.75\textwidth]{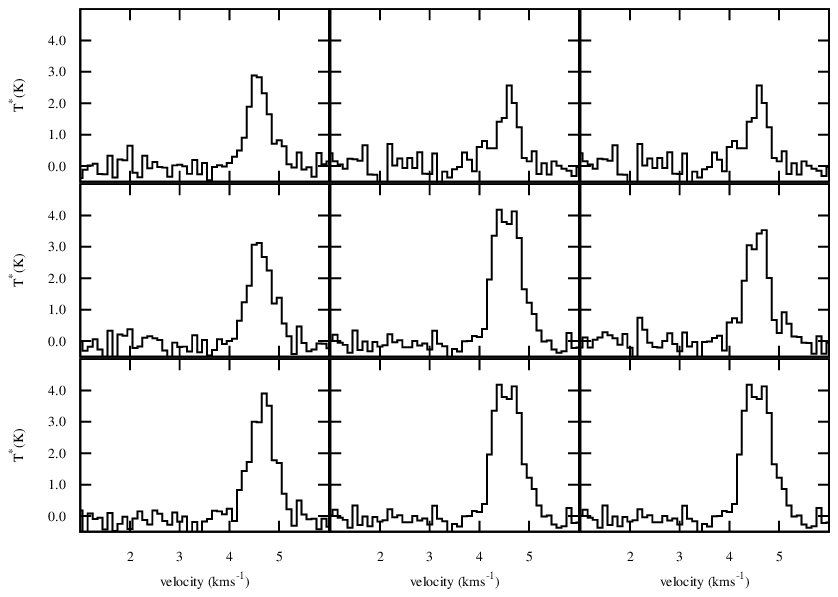}
\caption[Map of HCO$^{+}$ (J=4$\rightarrow$3) spectra in the core E-MM2a.]{Map of HCO$^{+}$ (J=4$\rightarrow$3) spectra in the core E-MM2a. The central grid of 3$\times$3 HARP pixels is shown here.}
\label{E-MM2a}
\end{figure*}

Figures~\ref{A-MM1} -- \ref{F-MM2a} show the HCO$^{+}$ (J=4$\rightarrow$3) spectral line profiles for the same cores seen in Figures~\ref{cores_bad} -- \ref{cores_rad}. These plots show the central 3$\times$3 HARP pixels for each core. These plots also show the N$_{2}$H$^{+}$ line centres from \citet{2007A&A...472..519A} as vertical dashed lines. These data are not available for all cores.

\clearpage

\clearpage

\clearpage

\clearpage

\clearpage

\clearpage

\clearpage

\clearpage

\clearpage

\clearpage

\clearpage

\clearpage

\clearpage

\clearpage

\clearpage

\clearpage

\clearpage

\clearpage

\clearpage

\clearpage

\clearpage

\clearpage

\clearpage

\clearpage

\clearpage

\begin{figure*}
\includegraphics[width=0.75\textwidth]{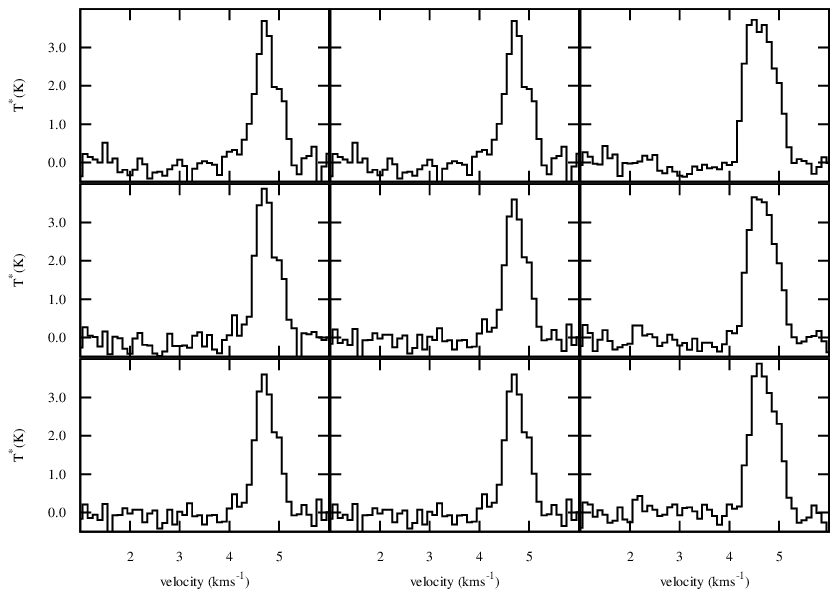}
\caption[Map of HCO$^{+}$ (J=4$\rightarrow$3) spectra in the core E-MM2b.]{Map of HCO$^{+}$ (J=4$\rightarrow$3) spectra in the core E-MM2b. The central grid of 3$\times$3 HARP pixels is shown here.}
\label{E-MM2b}
\end{figure*}

\clearpage

\begin{figure*}
\includegraphics[width=0.75\textwidth]{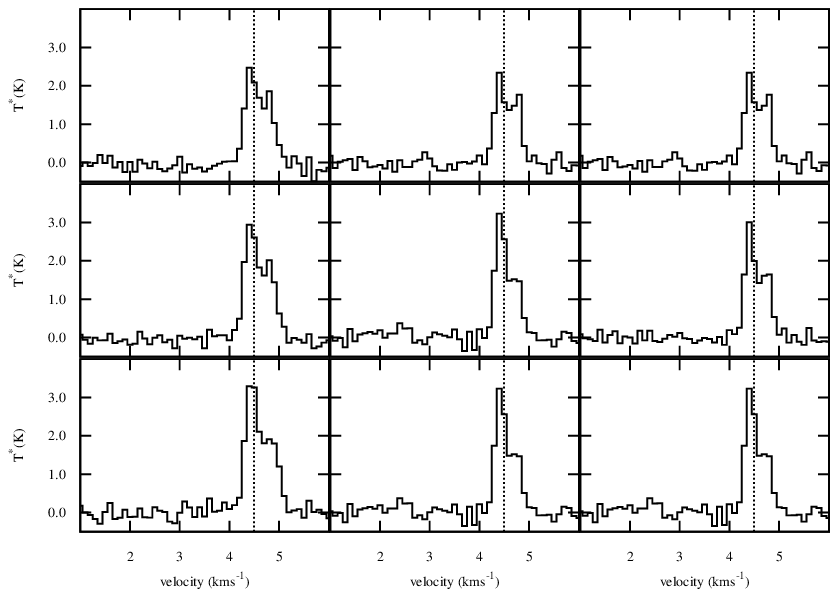}
\caption[Map of HCO$^{+}$ (J=4$\rightarrow$3) spectra in the core E-MM2d.]{Map of HCO$^{+}$ (J=4$\rightarrow$3) spectra in the core E-MM2d. The central grid of 3$\times$3 HARP pixels is shown here.}
\label{E-MM2d}
\end{figure*}

\begin{figure*}
\includegraphics[width=0.75\textwidth]{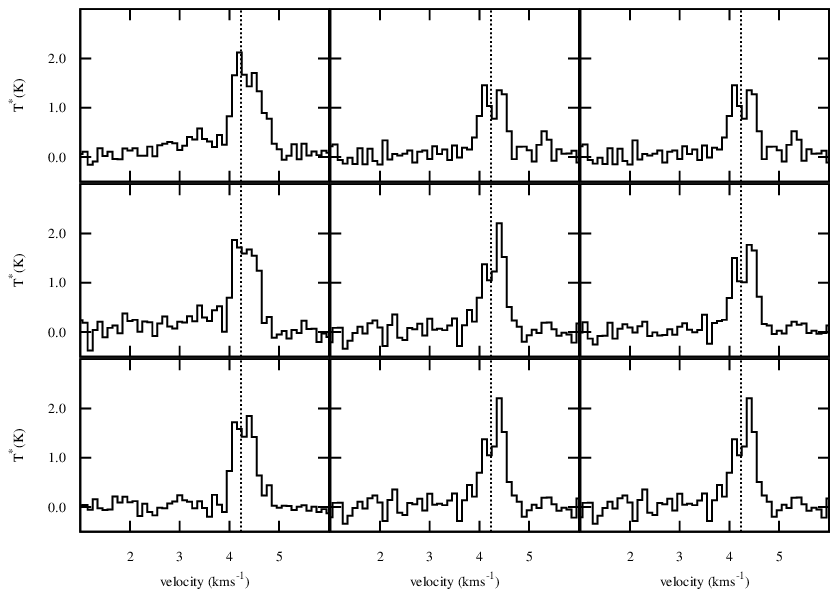}
\caption[Map of HCO$^{+}$ (J=4$\rightarrow$3) spectra in the core E-MM4.]{Map of HCO$^{+}$ (J=4$\rightarrow$3) spectra in the core E-MM4. The central grid of 3$\times$3 HARP pixels is shown here.}
\label{E-MM4}
\end{figure*}

\clearpage

\begin{figure*}
\includegraphics[width=0.75\textwidth]{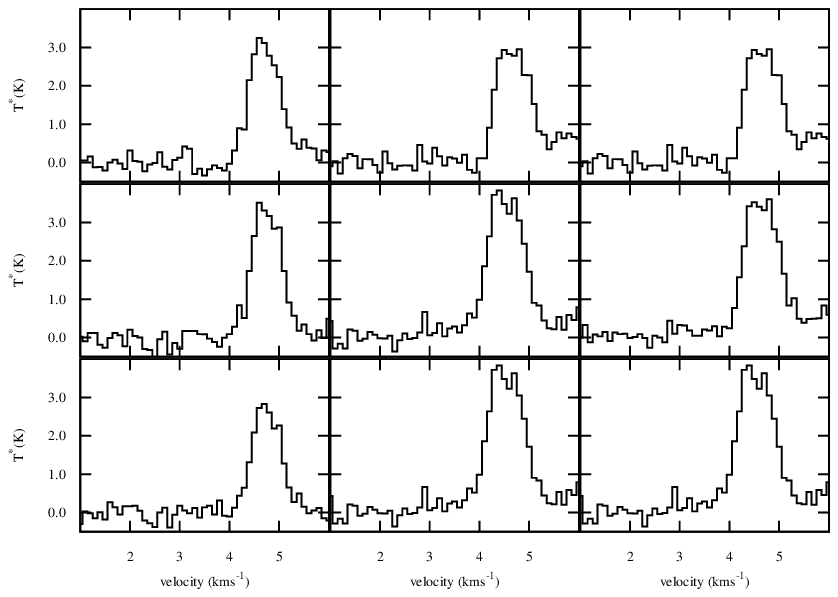}
\caption[Map of HCO$^{+}$ (J=4$\rightarrow$3) spectra in the core E-MM5.]{Map of HCO$^{+}$ (J=4$\rightarrow$3) spectra in the core E-MM5. The central grid of 3$\times$3 HARP pixels is shown here.}
\label{E-MM5}
\end{figure*}

\begin{figure*}
\includegraphics[width=0.75\textwidth]{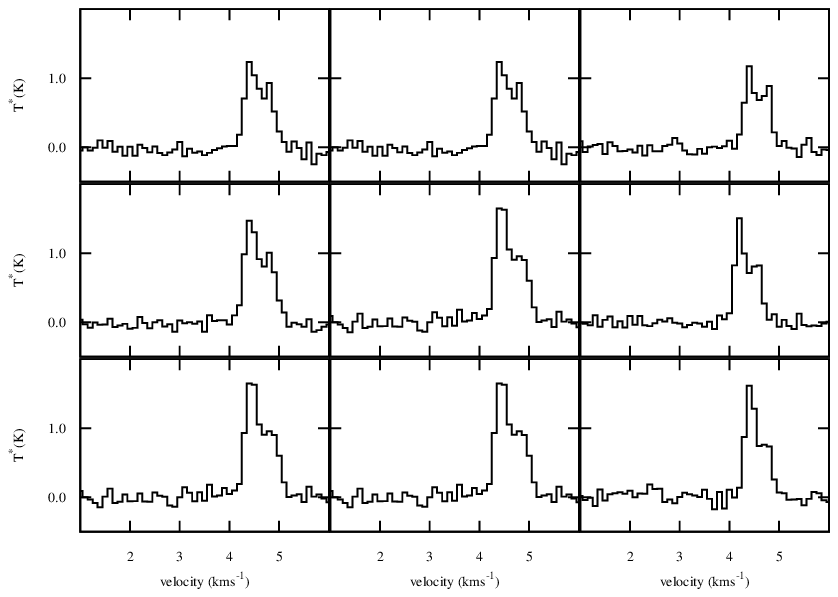}
\caption[Map of HCO$^{+}$ (J=4$\rightarrow$3) spectra in the core E-MM8.]{Map of HCO$^{+}$ (J=4$\rightarrow$3) spectra in the core E-MM8. The central grid of 3$\times$3 HARP pixels is shown here.}
\label{E-MM8}
\end{figure*}

\clearpage

\begin{figure*}
\includegraphics[width=0.75\textwidth]{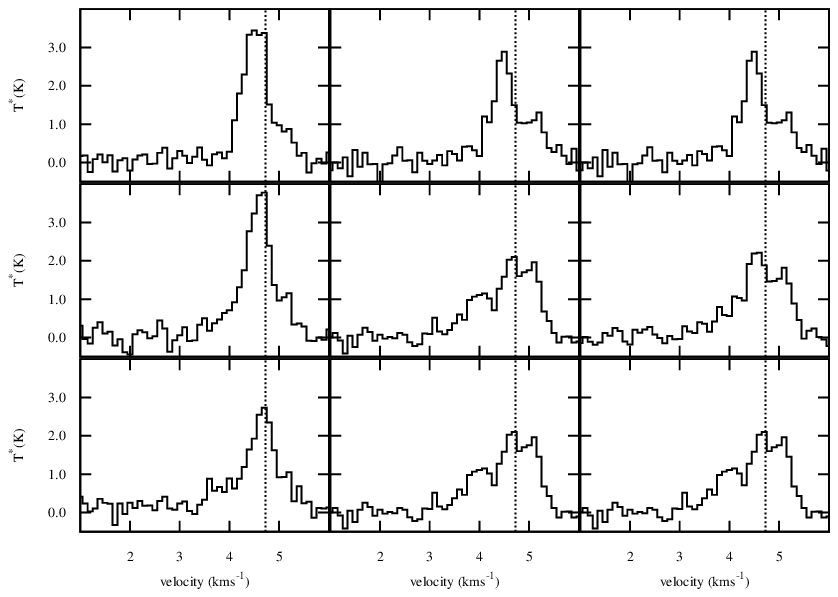}
\caption[Map of HCO$^{+}$ (J=4$\rightarrow$3) spectra in the core F-MM1.]{Map of HCO$^{+}$ (J=4$\rightarrow$3) spectra in the core F-MM1. The central grid of 3$\times$3 HARP pixels is shown here.}
\label{F-MM1}
\end{figure*}

\begin{figure*}
\includegraphics[width=0.75\textwidth]{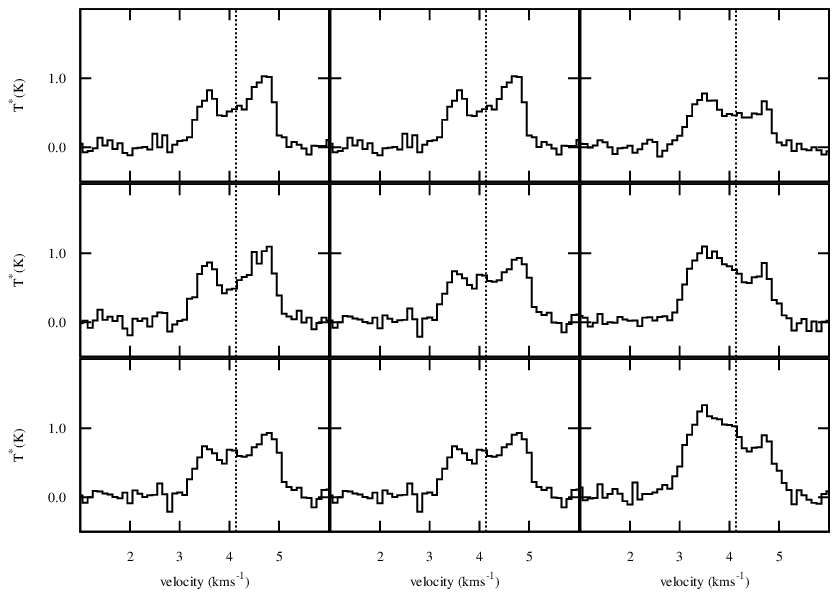}
\caption[Map of HCO$^{+}$ (J=4$\rightarrow$3) spectra in the core F-MM2a.]{Map of HCO$^{+}$ (J=4$\rightarrow$3) spectra in the core F-MM2a. The central grid of 3$\times$3 HARP pixels is shown here.}
\label{F-MM2a}
\end{figure*}

\end{document}